\PassOptionsToPackage{usenames,dvipsnames,table}{xcolor}
\documentclass[twocolumn,aps,prd,longbibliography]{revtex4-1}
\usepackage{graphicx,epsfig,epstopdf}
\usepackage{amsmath,amssymb,latexsym,bm}
\usepackage[english]{babel}
\usepackage{wrapfig}
\usepackage{animate}
\usepackage{parskip}
\usepackage{times}
\usepackage[T1]{fontenc}
\usepackage{color, colortbl}
\usepackage{tikz}
\usepackage[usenames, dvipsnames]{xcolor}
\usetikzlibrary{arrows,shapes}
\usepackage{url}
\usepackage{hyphenat}
\usepackage{makecell}
\usepackage[Conny]{fncychap}
\usepackage{float}
\usepackage{lipsum}
\usepackage{mathptmx}
\usepackage{mathtools}
\usepackage[T1]{fontenc}
\usepackage[%
  colorlinks=true,
  urlcolor=blue,
  linkcolor=blue,
  citecolor=blue
]{hyperref}
\DeclareUnicodeCharacter{2212}{-}
\begin{document}
\title{On the microstructure of higher-dimensional Reissner-Nordstr\"om black holes in quantum regime}


\author{Syed Masood A. S. Bukhari $^{1}$} \email{masood@zju.edu.cn}
\email{quantummind137@gmail.com}
\author{Behnam Pourhassan$^{2}$} \email{b.pourhassan@du.ac.ir}
\author{Houcine Aounallah$^{4}$} \email{houcine.aounallah@univ-tebessa.dz}
\author{Li-Gang Wang$^{1,3}$} \email{lgwang@zju.edu.cn}

\affiliation{$^{1}$ School of Physics, Zhejiang University, Hangzhou 310027, China\\
$^{2}$ School of Physics, Damghan University, Damghan, 3671641167, Iran\\
$^3$ Canadian Quantum Research Center 204-3002 32 Ave Vernon, British Columbia V1T 2L7 Canada\\
$^{4}$ Department of Science and Technology, Echahid Cheikh Larbi Tebessi
University - Tebessa, Algeria}
\date{\today}

\setlength{\parskip}{0cm}
    \setlength{\parindent}{1em}
\begin{abstract}
Thermodynamic Riemannian geometry provides great insights into the microscopic structure of black holes (BHs). One such example is the Ruppeiner geometry which is the metric space comprising the second derivatives of entropy with respect to other extensive variables of the system.  Reissner-Nordstr\"om black holes (RNBHs) are known to be endowed with a flat Ruppeiner geometry for all higher spacetime dimensions. However this holds true if one invokes classical gravity where the semi-classical Bekenstein-Hawking entropy best describes the thermodynamics of the system. If the much deeper quantum gravity and string   theories entail modifications to  BH entropy, this prompts the question whether the Ruppeiner flatness associated with higher dimensional RNBHs still persists. We investigate this problem by considering  non-perturbative (exponential) and perturbative (logarithmic) modifications to BH entropy of a $5$D RNBH. We find that while the case is so for larger (classical) geometries, the situation  is radically altered for smaller (quantum) geometries.  
Namely, we show surprising emergence of multiple phase transitions that depend on the choice of extent of corrections to BH entropy and charge. Our consideration involves differentiated extremal and non-extremal geometric  scales corresponding to the validity regime of corrections to entropy.
 More emphasis is laid on the exponential case as the contributions become highly non-trivial on small scales. An essential critical mass scale arises in this case that marks the onset of these phase transitions while the BH diminishes in size via Hawking evaporation. We contend that this critical value of mass perhaps best translates as the epoch of a classical to quantum BH phase transition.  
\end{abstract}
\date{\today}
\maketitle
\setlength{\parskip}{0cm}
    \setlength{\parindent}{1em}

\section{Introduction}
\label{sec:intro}
The powerful principle in Boltzmann's parlance: \textit{``If you can heat it, it has microscopic structure''} \cite{Wei:2015iwa}, has proven so instrumental in understanding the microstructure of physical systems. Ever since the ground-breaking works by Bekenstein \cite{Bekenstein:1973ur} and  Hawking \cite{Hawking:1974rv,
Hawking:1975vcx}, the study of black hole (BH) thermodynamics is thriving  as one of the major paradigms of modern physics. One of main lessons due to this is the fact that entropy of a BH, $S_{BH}=A/(4\ell_{p}^2)$, where $A$ is horizon area and $\ell_{p}$ is the Planck length, scales with its  area than volume, and this observation lies at the heart of \textit{holographic principle}  \cite{Susskind:1994vu,
Bousso:2002hon}.  This  relation quantifies the amount of entropy to be associated with a BH as a thermodynamic system as perceived by an external observer, providing a basis for conceiving BH microstructures.

As regards the final fate of BH shrinking via Hawking evaporation, one is forced to consider the quantum structure of spacetime geometry. Almost all known theories of quantum gravity necessitate the existence of a minimal length (often characterized by Planck length $\ell_{p}$) where classical geometry is plagued by quantum fluctuations \cite{Hossenfelder:2012jw}.  This entails radical consequences for entropy-area relation for a BH as it approaches  quantum regime \cite{Mann:1997hm,Upadhyay:2018vfu,Pourhassan:2019coq}, including holographic principle \cite{Bak:1998vj}. What happens is the modification of classical Bekenstein-Hawking relation,  and this obliquely counts as  deciphering the microscopic origin of BH entropy \cite{Strominger:1996sh}.

Numerous studies have  elucidated the way one  accounts for these modifications via different approaches, and interestingly these corrections enter the scenario   either through a perturbative or a non-perturbative framework. Perturbative methods include the microstate counting in string theory and loop quantum gravity \cite{Rovelli:1996dv,  PhysRevLett.80.904, PhysRevLett.84.5255,Rovelli:1996dv,Dabholkar:2008zy,Mandal:2010cj}, generally manifesting as logarithmic corrections, while  non-perturbative methods feature as exponential corrections \cite{Ashtekar:1991hf,Ghosh:2012jf, Dabholkar:2014ema, Chatterjee:2020iuf}. A prominent method to incorporate non-perturbative terms is by employing AdS/CFT correspondence \cite{Maldacena:1997re} and  using Kloosterman sums for massless supergravity fields near the horizon \cite{Murthy:2009dq, Dabholkar:2011ec, Dabholkar:2014ema}. For a large BH, all these corrections are suppressed and hence can be ignored, implying that Bekenstein-Hawking relation suffices to discuss the thermodynamic behavior of horizon. On contrary, for a smaller black hole, where quantum fluctuations become relevant, the logarithmic and other expansion terms also contribute, however still in a perturbative manner.  The most interesting situation arises  from the exponential term which dominates non-perturbatively and dramatically changes the physics around this regime. A considerable volume of literature is devoted to both perturbative and non-perturbatve corrections  in different contexts. For example, using holographic arguments based on AdS/CFT duality, quantum corrections to BH entropy have been computed in Refs. \cite{Hemming:2007yq, Gregory:2008br, Rocha:2008fe, Saraswat:2019npa}, including extremal geometries of Reissner-Nordstr\"{o}m 
\cite{Mann:1997hm} and rotating BHs \cite{Sen:2012cj}. The entropy of a conformal field theory can be obtained using Cardy's formula, and this approach has been used in Ref. \cite{Govindarajan:2001ee} to compute a leading order (logarithmic) correction to a Ba\~nados-Teitelboim-Zanelli (BTZ) black hole.  In Ref. \cite{Birmingham:2000xd}, authors investigate and analyze the sub-leading correction terms to Bekenstein-Hawking relation within conformal field theory.  For further insight into the problem, and many diverse aspects of (non-equilibrium) quantum thermodynamics of BHs, we refer the reader to the Refs. \cite{Pourhassan:2017qxi, Pourhassan:2017qhq, Pourhassan:2019coq, Pourhassan:2020bzu, Upadhyay:2019hyw, Ghaffarnejad:2022aqe, Biswas:2021gps,Pourhassan:2022sfk, Pourhassan:2022opb, Aounallah:2022rfo, Aounallah:2023pdf} for a comprehensive look.

The above formulations are all built on the notion of an existing gravitational system with a well defined geometry. However, it is quite possible that if one starts from a thermodynamic footing viz. entropy and Clausius relation, the result is an emergent geometry.   A  seminal paper by Jacobson \cite{PhysRevLett.75.1260} laid the foundation for a thermodynamic viewpoint on Einstein gravity. The central idea is  the ubiquity of the Clausius relation, $\delta Q=TdS$, where $\delta Q$ is the matter-energy flux crossing a local Rindler horizon with an associated Unruh temperature $T$, supplemented by entropy-area correspondence. Consequently, the equations of general relativity emerge as a thermodynamic equation of state in a natural way. Since then the original idea has been generalized in many ways and led to many new ideas and great insights  \cite{PhysRevLett.96.121301, Padmanabhan_2010,PhysRevD.81.024016, PhysRevD.85.064017, PhysRevLett.116.201101, PhysRevD.98.026018, Alonso-Serrano:2020dcz}. For example,  some higher-curvature gravity models have been shown to possess intriguing thermodynamic interpretation leading to an \textit{emergent gravity paradigm} \cite{Padmanabhan:2009jb,Padmanabhan:2009vy,Padmanabhan:2014jta}. It is noteworthy that logarithmic corrections to entropy also arise due to  thermal fluctuations around an equilibrium configuration without any need for an underlying quantum gravity theory \cite{Das:2001ic}. However, an extension of Jacobson formalism  relates thermal fluctuations to quantum geometry fluctuations \cite{Faizal:2017drd}.

In light of the realization that quantum gravity predicts corrections to classical thermodynamic variables, it is reasonable to assume that this thermodynamics which holds in both classical and quantum domains of spacetime geometry might be able to suggest the modifications to gravitational dynamics of Einstein equations. This is the very principle underlying the motivation of this work.  However, the present work  only investigates the modified thermodynamics including the phase transitions  based on (non-)perturbative corrections  to BH entropy \cite{Dabholkar:2011ec, Dabholkar:2014ema, Chatterjee:2020iuf}, without going to compute corrections to spacetime geometry.
We discuss the consequences of these non-perturbative (exponential) and perturbative (logarithmic) corrections to a five-dimensional ($5$D) Reissner-Nordstr\"{o}m BH (RNBH). We address the question of thermodynamic (un-)stability via  the information geometric approach.\\
\indent It is well-known that usual picture of thermodynamics of BHs including the one via information geometric approach builds on the idea of fluctuations around equilibrium description. However, this equilibrium description breaks down when BHs attain quantum size, possibly around Planck scale \cite{Pourhassan:2022sfk}. Thus one needs a non-equilibrium description for BH thermodynamics \cite{Pourhassan:2022opb,Iso:2011gb}. Thus our study of thermodynamic phase trasitions should be taken in this sense.\\
\indent It is crucial to mention here our choice of higher dimensions in this work leads to some rich behaviour for phase structure compared to the $4$D counterpart, as we shall see later. For $4$D case, earlier study has already been performed regarding modified thermodynamics under the impact of exponential corrections \cite{Pourhassan:2020yei}. However, as thermodynamic geometry offers more potent means to explore the phase structure, we will also analyze Ruppeiner geometry for $4$D RNBH and compare it with the $5$D case. Note that as we mentioned earlier our prime focus is on exponential case.

The working principle we adhere to here involves a canonical ensemble-type approach having a constant charge with minuscule gravitational contributions. This helps us to differentiate classical and quantum regimes of geometry corresponding to non-extremal and extremal scenarios, respectively. Having carefully defined this, the regime of application for quantum corrections to BH entropy follows naturally.

The paper is organized as follows. In Sec.\ref{sec:concepts}, we review the geometry of higher-dimensional RNBH and the modifications to BH entropy. Sec. \ref{sec:thermstability} details the stability analysis of $5$D RNBH based on modified heat capacity of a BH system. Sec. \ref{sec:Ruppgeometry} provides a discussion of Ruppeiner approach to thermodynamic geometry and hence we compute the associated curvature scalar for our system. The conclusion is drawn in Sec. \ref{sec:conclusion}.

\section{\label{sec:concepts}
Conceptual aspects: Higher-dimensional Reissner-Nordstr\"{o}m geometry and corrections to entropy}
Theories beyond general relativity, in addition to many others, include a class of higher dimensional models of gravity that hold great scope from mathematical and physical point of view. The initiation is rooted in the ideas by  Kaluza and Klein \cite{Kaluza:1921tu, 1926ZPhy...37..895K} as a way to unify  electromagnetism with gravity and currently occupies a special position in string theories \cite{Chan:2000ms}. The situation is akin to quantum field theory where one can chose an arbitrary field content beyond the existing boundaries of  Standard Model, shedding light on many general features of quantum fields. The hope here is a D-dimensional extension of general relativity could lead to valuable insights into the theory and especially into one of its robust predictions, the BHs \cite{Emparan:2008eg}. As is well known, the hallmarks of  $4$D BHs comprise   spherical topology, dynamical stability, uniqueness,  and satisfying a set of basic rules-the laws of BH mechanics. A growing understanding suggests that gravity offers  much richer physics in D$>4$ dimensions, as evinced by the discovery of dynamical instabilities in extended horizon geometries \cite{Gregory:1993vy}, and BHs endowed with non-spherical topologies not generally identified with uniqueness-a trait otherwise associated with their $4$D counterparts \cite{PhysRevLett.88.101101}. A more fascinating result  links higher dimensional BHs to fluid dynamics in the so-called \textit{blackfold} approach \cite{Emparan:2009at}.  In addition, it has been shown that the behaviour of higher dimensional BH thermodynamics is affected in an energy-dependent background geometry \cite{Aounallah:2020yjo}.

The inclusion of extra dimensions in BHs dates back to Tangherlini \cite{Tangherlini:1963bw}, who formulated  a D-dimensional solution for Schwarzschild  and  RNBH. The action is given by \cite{Destounis:2019hca}
\begin{eqnarray}
 S=-\frac{1}{16\pi G_{D}}\int d^Dx\sqrt{-g}(R-F^2),
\end{eqnarray}
where $G_{D}$ is D-dimensional  Newton's gravitational constant, $g$ is the determinant of metric tensor $g_{\mu\nu}$, $R$ represents Ricci scalar,  and $F^2$ the electromagnetic Lagrangian, which yields a D-dimensional RN spacetime. The resultant metric, which represents a static and spherically symmetric spacetime, is given by 
\begin{equation}\label{metric}
ds^{2}=-f(r)dt^{2}+\frac{dr^{2}}{f(r)}+r^{2}d\Omega_{D-2}^{2},
\end{equation}
where $d\Omega_{D-2}$ is the metric of unit $(D-2)$-sphere. Here the metric function $f(r)$ reads 
\begin{equation}
f(r)=1-\frac{2\mu}{r^{D-3}}+\frac{q^{2}}{r^{2(D-3)}}.
\end{equation}
The parameters $\mu$ and $q$ are some constants that help us to define Arnowit-Deser-Misner (ADM) mass $M$  and electric charge $Q$ of BH as
\begin{eqnarray}
M=\left(\frac{D-2}{16\pi}W_{D-2}\right)\mu,\ 
Q=\left(\frac{\sqrt{2(D-2)(D-3)}}{8\pi}W_{D-2}\right)q,
\end{eqnarray}
with $ W_{D}=\left(2\pi^{\frac{D+1}{2}}\right)/\Gamma(\frac{D+1}{2})$. We deal with D$=5$ case, hence the above parameters read 
\begin{equation}\label{fr}
f(r)=1-\frac{2\mu}{r^{2}}+\frac{q^{2}}{r^{4}}, \ \ M=\frac{3\pi}{4}\mu, \ \ Q=\frac{\sqrt{3\pi}}{4}q.
\end{equation}
Since there is a linear mapping between $(\mu,q)$ and $(M,Q)$ respectively, we can safely  treat  $M$ and $Q$ as our mass and charge throughout the work.  
For the non-extremal case $M> Q$, the zeros of $f(r)$ give two horizons located at    
\begin{equation}\label{root}
r_{\pm}=\sqrt{M\pm\sqrt{M^{2}-Q^{2}}},
\end{equation}
where $r_{+}$ is the BH event horizon and $r_{-}$ is the inner Cauchy horizon. The temperature is defined from the metric function $f(r)$ as
\begin{eqnarray}
T&=\frac{1}{4\pi}\left(\frac{df\left(r\right)}{dr}\right)_{r=r_{+}}&=\frac{4M}{r^3}-\frac{4Q^2}{r^5},
\end{eqnarray}
where we dropped the constant factor $1/4\pi$ for convenience.
We substitute $r_{+}$ from eq.(\ref{root}) in the above equation to obtain $T$ in terms of $M$ and $Q$. Hence we have
\begin{eqnarray}\label{temp1}
 T=\frac{4 M}{\left(\sqrt{M^2-Q^2}+M\right)^{3/2}}-\frac{4 Q^2}{\left(\sqrt{M^2-Q^2}+M\right)^{5/2}}.
\end{eqnarray}

\begin{figure*}[htb!]
\centering
\includegraphics[height=6.7cm,width=8cm]{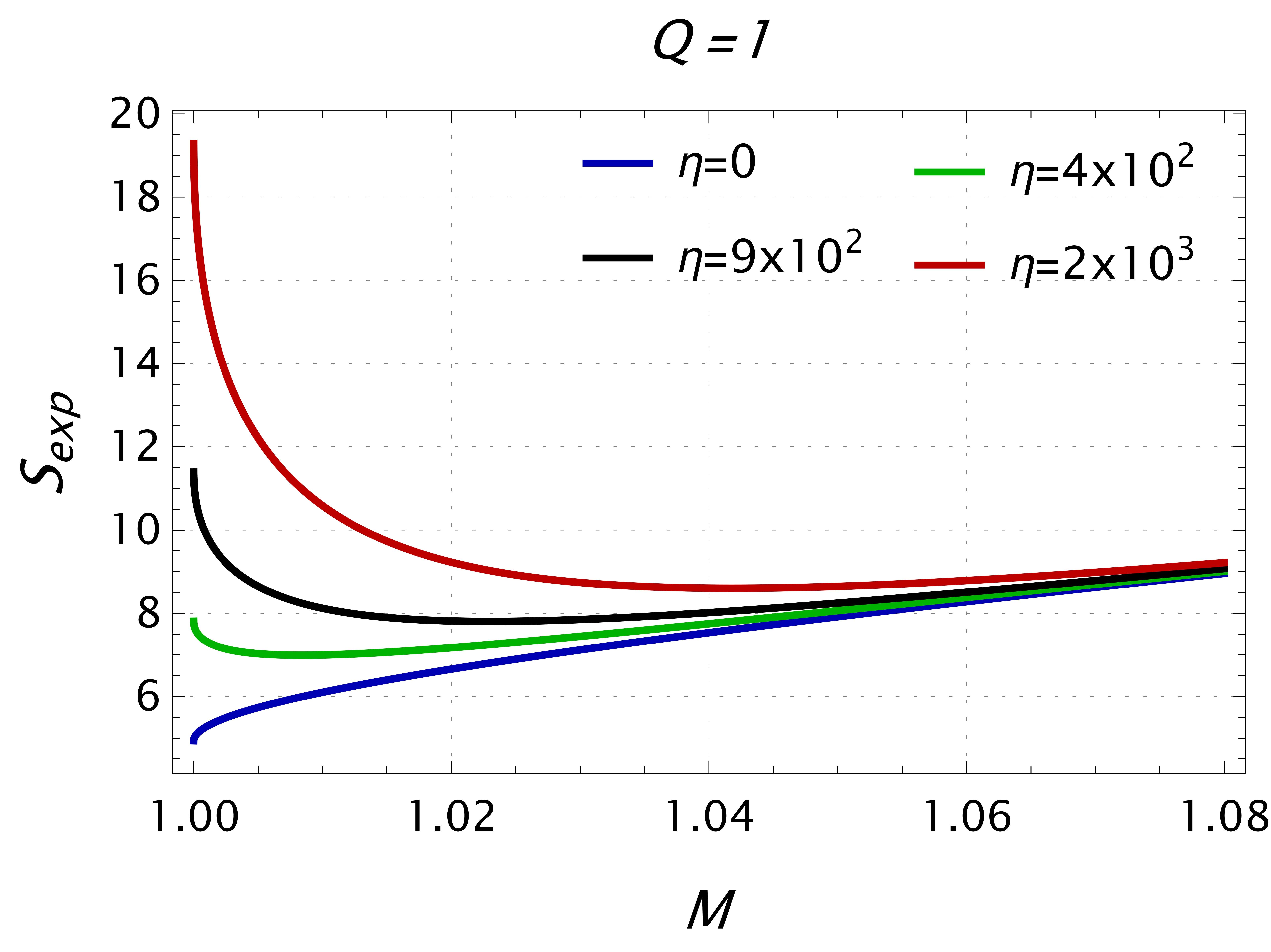}
\caption{ Entropy $S_{exp}$ variation with mass $M$ for a $5$D RNBH with non-perturbative corrections. The correction parameter $\eta$ greatly controls  entropy behavior as the BH shrinks to smaller sizes. Note that at extremal limit,  $S_{exp}$ possesses a large (remnant) value and \textit{does not} diverge.}
\label{entropy}
\end{figure*}

\begin{figure*}[htb!]
\centering
\includegraphics[height=6.7cm,width=0.484\textwidth]{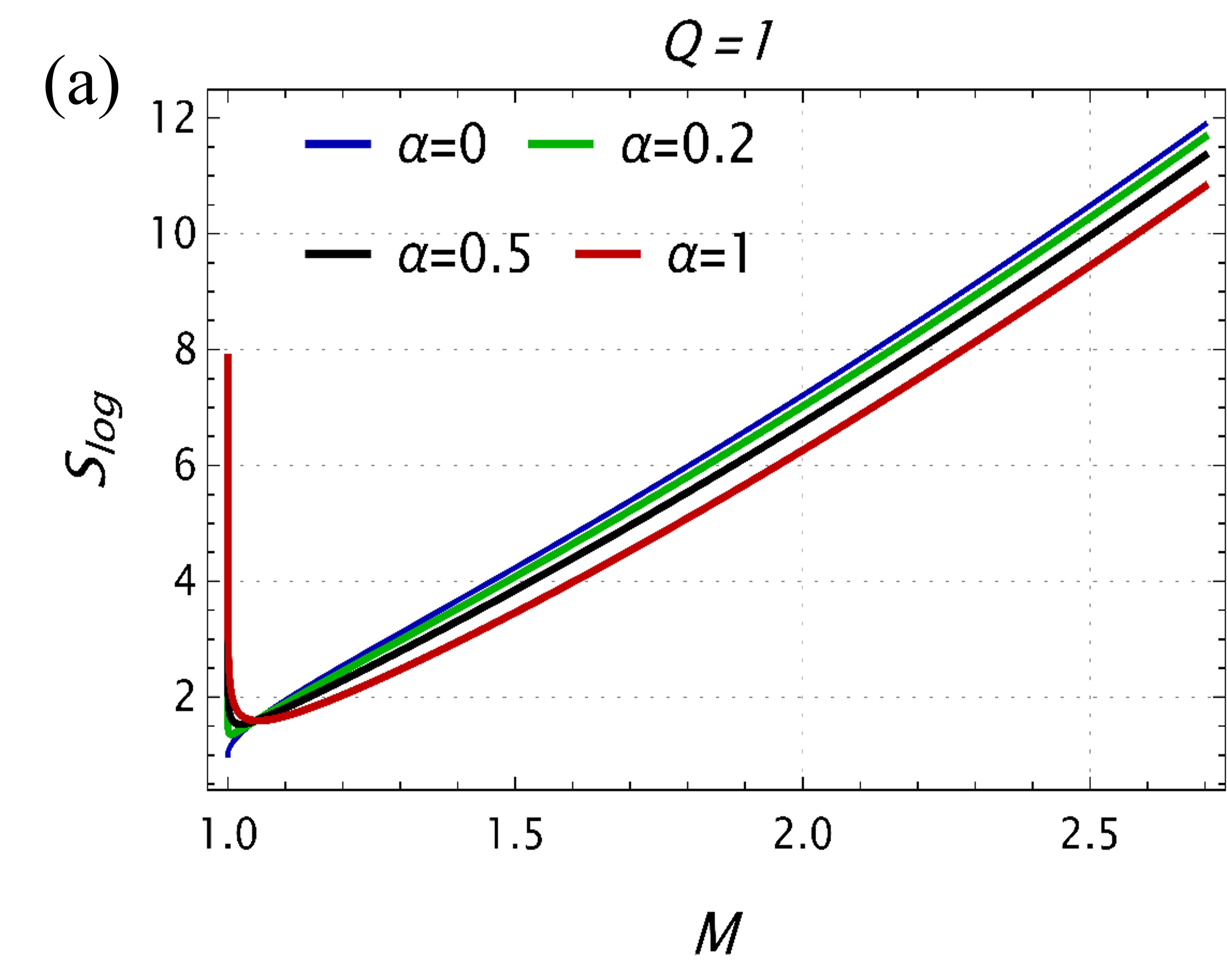}\ \ 
\includegraphics[height=6.7cm,width=0.484\textwidth]{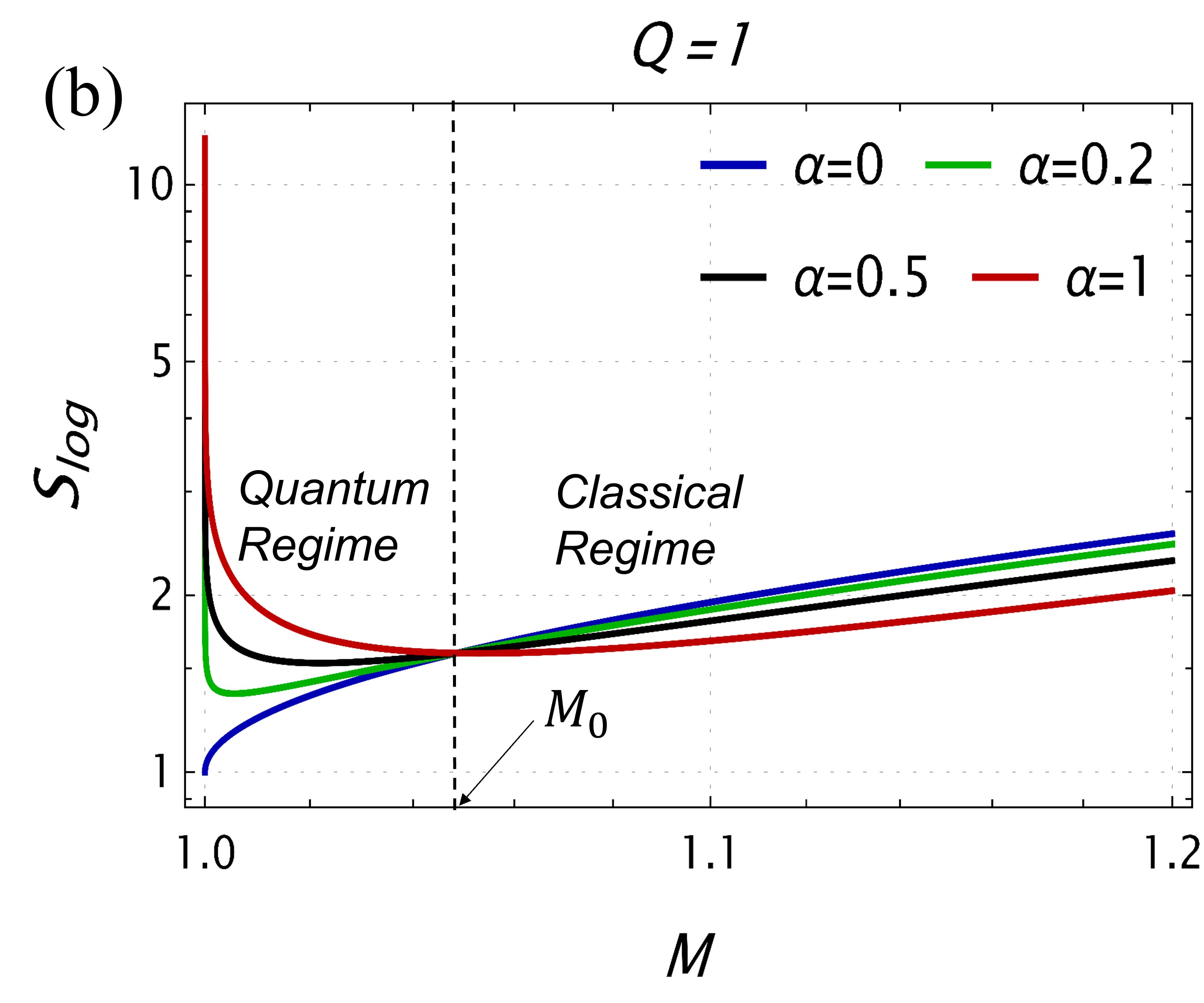}
\caption{(a) Entropy $S_{log}$ variation with mass $M$ for logarithmic corrections, and (b) short-distance behaviour of $S_{log}$ on log-log scale.}
\label{Slog}
\end{figure*}

As regards the microscopic origin of BH entropy that invokes a full theory of quantum gravity, the first of its kind was reported in string theory framework \cite{Strominger:1996sh}, and interestingly string theory rests on extra dimensions. As pointed out earlier, the modifications to original Bekenstein-Hawking entropy become appreciable when the hole size approaches quantum gravity scale, any underlying quantum gravity theory that doesn't produce the original entropy relation in leading order is surely incorrect. Even though quantum BH geometry is itself a model-dependent approach and one has to rather start from quantizing the gravitational action that may prove a daunting task if accounting for non-perturbative corrections \cite{Pourhassan:2020all}. However,  the quantum corrections to entropy is another powerful way to study the end stage of BH as the size decreases. The primary impetus comes from Jacobson formalism \cite{PhysRevLett.75.1260} and non-equilibrium thermodynamics \cite{Das:2001ic}, and relating thermal fluctuations to geometry \cite{Faizal:2017drd}.  The class of perturbative quantum corrections to BH entropy usually assume the general form \cite{PhysRevLett.80.904, PhysRevLett.84.5255,Pourhassan:2017qhq,Pourhassan:2019coq}
\begin{eqnarray}
 S_{p}=\alpha\ln{\left(\frac{A}{4\ell_{p}^2}\right)}+\frac{ 4\beta \ell_{p}^2}{A}+...,
\end{eqnarray}
where $A=4\pi r_{+}^2$ is the area of event horizon,  and $\alpha$ and $\eta$ are constants. The non-perturbative corrections are of the following form \cite{Dabholkar:2011ec, Dabholkar:2014ema, Chatterjee:2020iuf}:
\begin{eqnarray}
 S_{np}=\eta e^{-A/4\ell_{p}^2}.
\end{eqnarray}
The total BH entropy is the sum of original entropy $S_{0}=A/4\ell_{p}^2$, perturbative and non-perturbative terms,
\begin{align}\nonumber
 S_{BH}&=S_{0}+S_{p}+S_{np}\\
 &=\frac{A}{4\ell_{p}^2}+\alpha\ln{\left(\frac{A}{4\ell_{p}^2}\right)}+\frac{\beta 4\ell_{p}^2}{A}+\eta e^{-A/4\ell_{p}^2}+...
\end{align}
It is important to note here that the above functional form is valid for all $4$D BHs, and the parameters $(\alpha,\beta,\eta)$ in the above equation signify the scale at which the corrections become relevant and can be obtained by a quantum corrected action which yields the required $4$D RNBH. For ordinary $5$D RNBH, the entropy is given by
\begin{eqnarray}
 S_{0}&=\frac{1}{2}\pi^2r_{+}^3,
\end{eqnarray}
 with the horizon area $2\pi^2r_{+}^3$.
Since the corrections apply to all BHs in general, we conjecture that a $5$D RNBH also receives the corrections.
Our focus here is to examine the effect of exponential term given by \cite{Chatterjee:2020iuf}
\begin{align}\nonumber
 S_{exp} &=\frac{1}{2}\pi^2r_{+}^3+\eta \exp{\left(-\frac{1}{2}\pi^2r_{+}^3\right)}\\
 \label{ourentropy}
 &=\frac{1}{2} \pi ^2 \left(\sqrt{M^2-Q^2}+M\right)^{3/2}+\eta e^{-\frac{1}{2} \pi ^2 \left(\sqrt{M^2-Q^2}+M\right)^{3/2}},
 \end{align}
 and the logarithmic term given by \cite{Das:2001ic}
 \begin{align}\nonumber
S_{log}&= S_{0}-\frac{1}{2}\alpha\log{\left(S_{0}T^2\right)}\\
 &=\left(\sqrt{M^2-Q^2}+M\right)^{3/2}-\frac{1}{2} \alpha  \log \Bigg[\left(\sqrt{M^2-Q^2}+M\right)^{3/2}  \\
\label{SLOG} & \times \left(\frac{4 M}{\left(\sqrt{M^2-Q^2}+M\right)^{3/2}}-\frac{4
   Q^2}{\left(\sqrt{M^2-Q^2}+M\right)^{5/2}}\right)^2\Bigg],
\end{align}
where the parameters $\eta$ and $\alpha$ characterize the extent of exponential and logarithmic corrections, respectively. The range of $\eta$ can be taken as far as the original Bekenstein-Hawking contribution is dominant, i.e. the exponential terms are suppressed for bigger sizes. For $\alpha$, the parameter range is usually taken as $\alpha\in[0,1]$ \cite{PhysRevD.94.064006}. 
  We  plot entropy for exponential $(S_{exp})$ and logarithmic $(S_{log})$ corrections, respectively, in figures \ref{entropy} and \ref{Slog}.
  We use these relations to discuss the thermodynamic stability and phase transition for our BH system in next section. 

Figure \ref{entropy} shows that the entropy $S_{exp}$ does not vanish for $5$D RNBH as it  evaporates to smaller sizes in presence of non-perturbative corrections, and this signals the onset of quantum fluctuations (scaled by $\eta$), while agreeing with Bekenstein-Hawking contribution for larger sizes. For unperturbed BHs ($\eta=0$), we recover the full Bekenstein-Hawking entropy for $5$D case. 
For logarithmic corrections, entropy is plotted in figure \ref{Slog}. It is clear that due to log corrections, BH possess less entropy compared to the original one. However, it is seen that as extremal limit $(M=Q)$ is reached, there is a sudden rise in entropy. However, there is a no singularity, just as is the case with exponential corrections. 

One can easily appreciate from Fig. \ref{Slog} (a) that logarithmic corrections are universal for all BH sizes but contribute perturbatively. However, in quantum regime, they show significant impact as seen in figure \ref{Slog} (b). Note that the dashed boundary, related to a critical mass $M_{0}$, separates the role of $\alpha$ for classical and quantum domains of BH geometry. This would appropriately be treated as the point of classical to quantum phase transition. $M_{0}$ would be the value of BH mass such that the term inside the logarithm of Eq. (\ref{Slog}) is $1$, i.e. whenever 
\begin{widetext}
\begin{eqnarray*}
     \left(\frac{4 M}{\left(\sqrt{M^2-Q^2}+M\right)^{3/2}}-\frac{4
   Q^2}{\left(\sqrt{M^2-Q^2}+M\right)^{5/2}}\right)^2=\frac{1}{\left(\sqrt{M^2-Q^2}+M\right)^{3/2} }.
\end{eqnarray*}
    \end{widetext}
Setting $Q=1$, we numerically find the value of $M_{0}\approx 1.04861$.

As stated earlier, the scope of quantum corrections to  entropy is tied up to the scale of geometry. It is thus imperative to make our definitions clear.  As in the traditional approach, the usual picture of BH evaporation involves shrinkage of whole BH horizon radius $r_{+}$ (containing both $M$ and $Q$). It is corroborated by the fact that for  charged BHs, the charge to mass ratio evolves with time as the evaporation continues \cite{Hiscock:1990ex}, possibly  proceeding toward an extremal geometry. It is this epoch where our definition applies.  As we work in a canonical ensemble paradigm, there is no harm in  treating $Q$ as a constant parameter throughout the process, as if   $M$ dictates the size of our system.  However, the caveat is that $Q$ is extremely small so as to help clearly differentiate the classical-quantum split as $M$ decreases due to evaporation. Hence, following this logic, whenever $M>Q$ (non-extremal case), our BH geometry is classical, and as $M\rightarrow Q$ (extremal), it possesses  a quantum description. These two phases kind of coexist at the critical mass $M_{0}$, which we indicated above for logarithmic case. For exponential case, we will define it from heat capacity.

\section{\label{sec:thermstability} Thermodynamic Stability and phase transition: The role of           
 (non-) perturbative corrections}
\subsection{Is $5$D RNBH colder than its Schwarzschild counterpart?}
We answer this question by considering the temperature $T$ of a $5$D RNBH.
\begin{figure}[htbp]
\centering
\includegraphics[height=6.6cm,width=8cm]{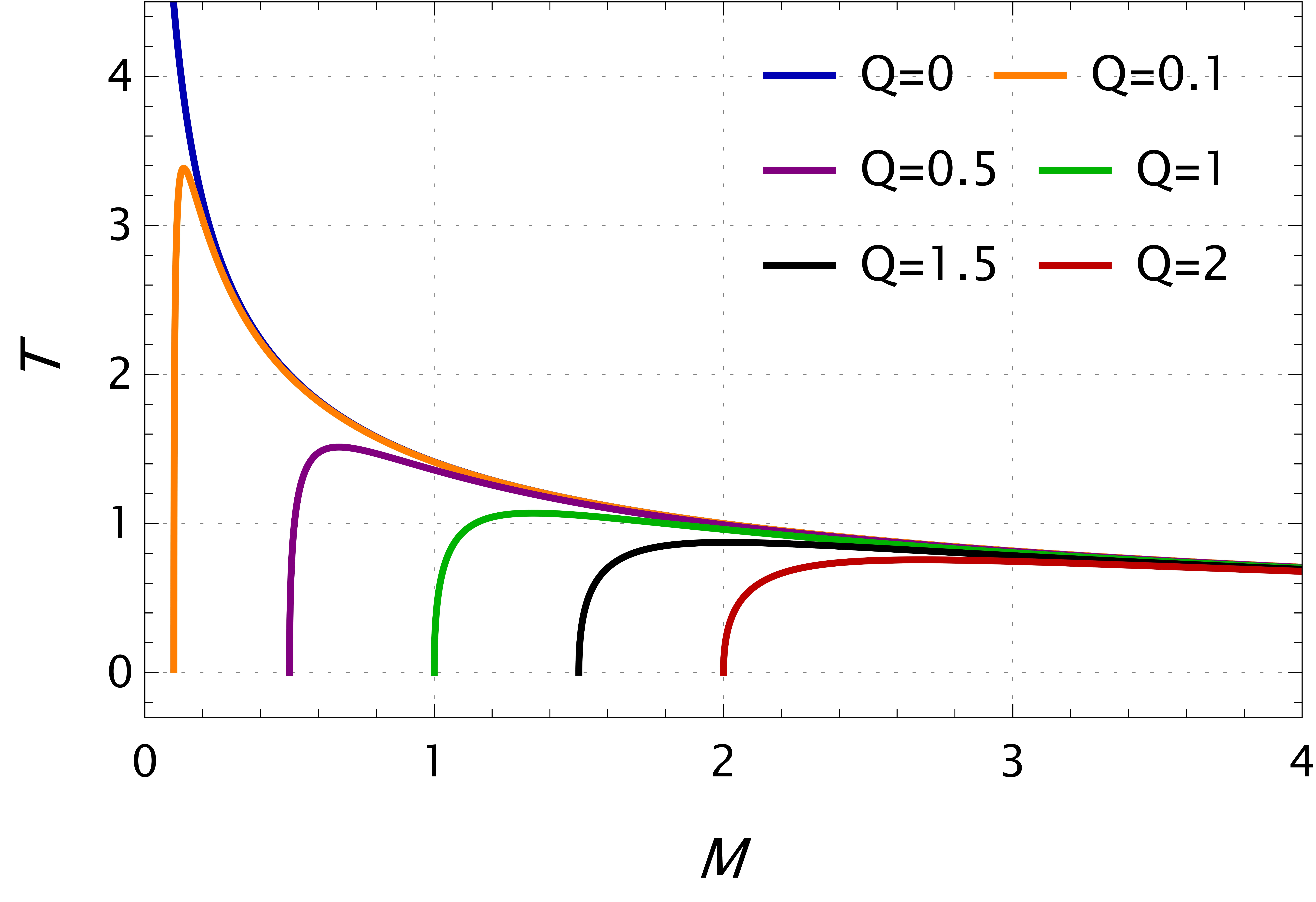}
\caption{\label{T1} Temperature $T$ vs BH mass $M$: the role of charge $Q$ is to reduce the temperature of RNBHs compared to Schwarzschild case.}
\end{figure}
From Eq. (\ref{temp1}), one can determine the behavior of temperature in terms of BH mass $M$ as the evaporation continues. By fixing $Q$, we plot temperature $T$.  It is well known that BHs radiate via Hawking radiation at $T$. From Fig. \ref{T1}, at first glance, it is evident   that temperature $T$ of our BH $(Q\neq 0)$ is less than its neutral $(Q=0)$ counterpart (Schwarzschild BH). Thus a charged BH is always colder than its neutral cousin. This in other words reflects the fact that a charged BH emits fewer neutral massless particles than an uncharged one. It is noteworthy that this result is quite known for D$=4$ spacetime dimensions \cite{Good:2020qsy}, and interestingly continues to hold even in extra-dimensional case. Also note that $T$ decays with increasing $M$ for both charged and uncharged cases due to competition between the two terms in eq. (\ref{temp1}). It is crucial to note here that we did not invoke the effect of  $\eta$ and $\alpha$ on $T$. This can be argued on the basis of simple realization that all possible perturbative and non-perturbative corrections to entropy, which are related to or inspired from AdS/CFT correspondence, can be expressed as functions of original entropy and temperature \cite{Mukherji:2002de, Lidsey:2002ah, Dehghani:2002jh, Das:2003fp}. In more explicit terms, what we mean here is that new modified entropy is expressible in terms of original $T$. For example, in Eq. (\ref{SLOG}), logarithmic corrections are expressible in terms of original  $T$, and these corrections find roots in AdS/CFT correspondence, as elucidated in the above works. This is further reflected from the fact that the  usual definition of the temperature stems from the metric function of the spacetime $(\propto df(r)/dr)$, and metric  represents  the geometry of spacetime.   As mentioned previously, the modifications only apply to the thermodynamics of the spacetime, and not the geometry, hence $T$ remains same as original one. We want to mention here though that this is an approximation taken in our case. In principle, it is possible to invoke quantum corrections to the geometry as well, which is however beyond the scope of our work.
\subsection{Remnant formation, phase structure and instabilities}
A  remnant is the leftover structure after  a BH ceases its Hawking evaporation. Technically speaking, a remnant is a localized late stage outcome of Hawking evaporation. This generally occurs in almost all quantum gravity and string  theories where spacetime structure approaches Planck scale, and  is more suitably expressed by quantum geometry. This discussion is central to the BH information paradox \cite{Chen:2014jwq}. Here, our motive is is to uncover the implications for thermodynamic behaviour of 5D RNBH.

We now turn our attention to the study of  thermodynamic stability and the conditions under which our system shows phase transition. This is possible by studying the variation of different thermodynamic  quantities or state functions in either canonical or grand canonical ensemble. Canonical ensemble theory assumes charge $Q$ to be a fixed parameter. Therefore, heat capacity, denoted by $C_{Q}$, dictates the stability conditions for such a system. The positivity of  $C_{Q}$ guarantees a stable phase and vice-versa, and  the divergence a phase transition \cite{1978RPPh...41.1313D, Chamblin:1999hg}. In particular, a vanishing and divergent heat capacity corresponds to \textit{first} and \textit{second order phase transitions}, respectively \cite{PhysRevD.91.124057}. It has been argued, however, that the phase transition analysis is more plausible by utilizing thermodynamic curvature rather than heat capacity \cite{Ruppeiner:2013yca}.  In general, $C_{Q}$ is defined as  
\vspace{-10pt}
\begin{eqnarray}\label{CQ}
C_{Q}=T\left(\frac{\partial S}{\partial T}\right)_{Q}=T\left(\frac{\partial S/\partial M}{\partial T/\partial M}\right)_{Q}.
\end{eqnarray}
\subsubsection{Exponential corrections}
\begin{figure*}
\includegraphics[height=6.7cm,width=0.484\textwidth]{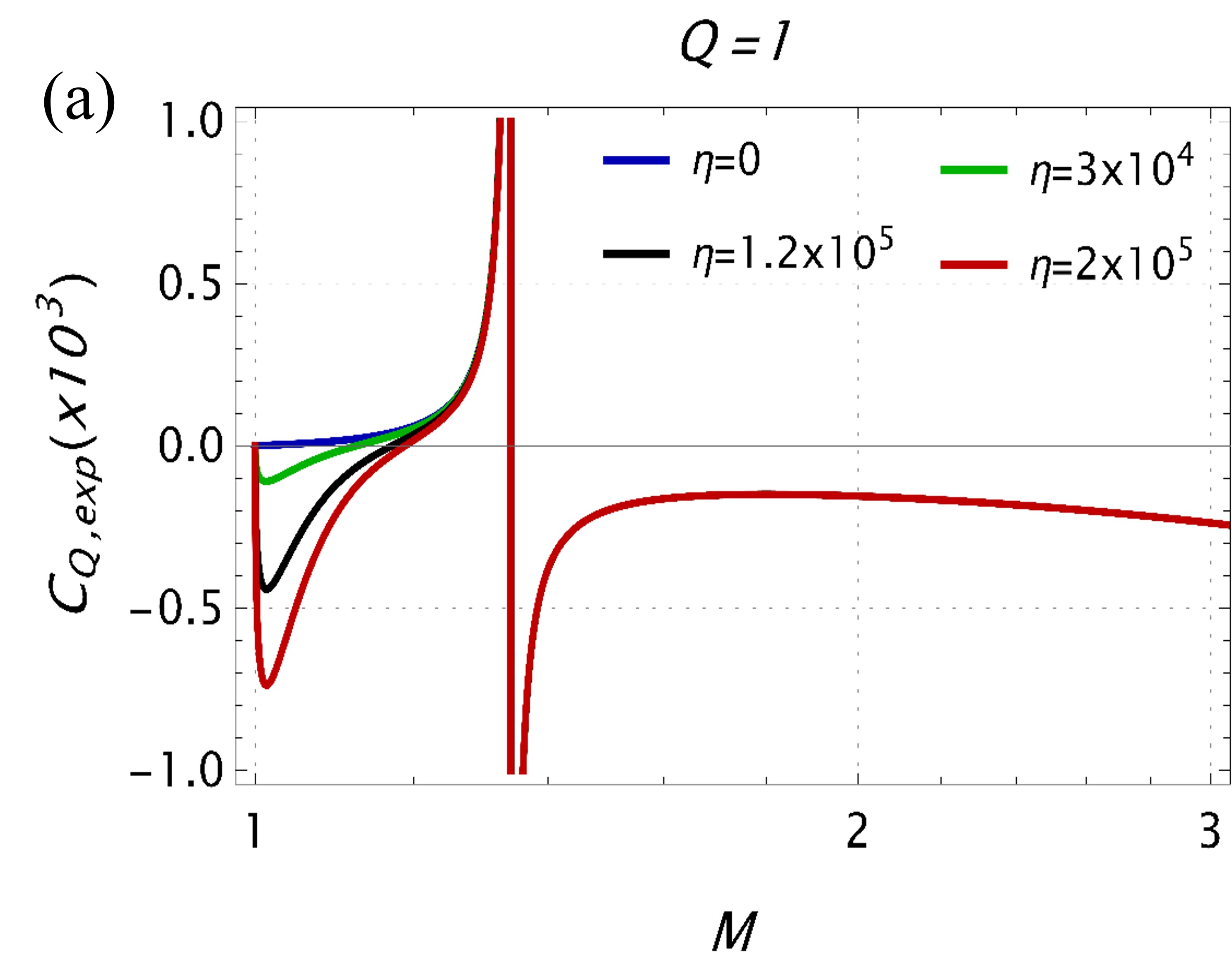}\ \ 
\includegraphics[height=6.7cm,width=0.485\textwidth]{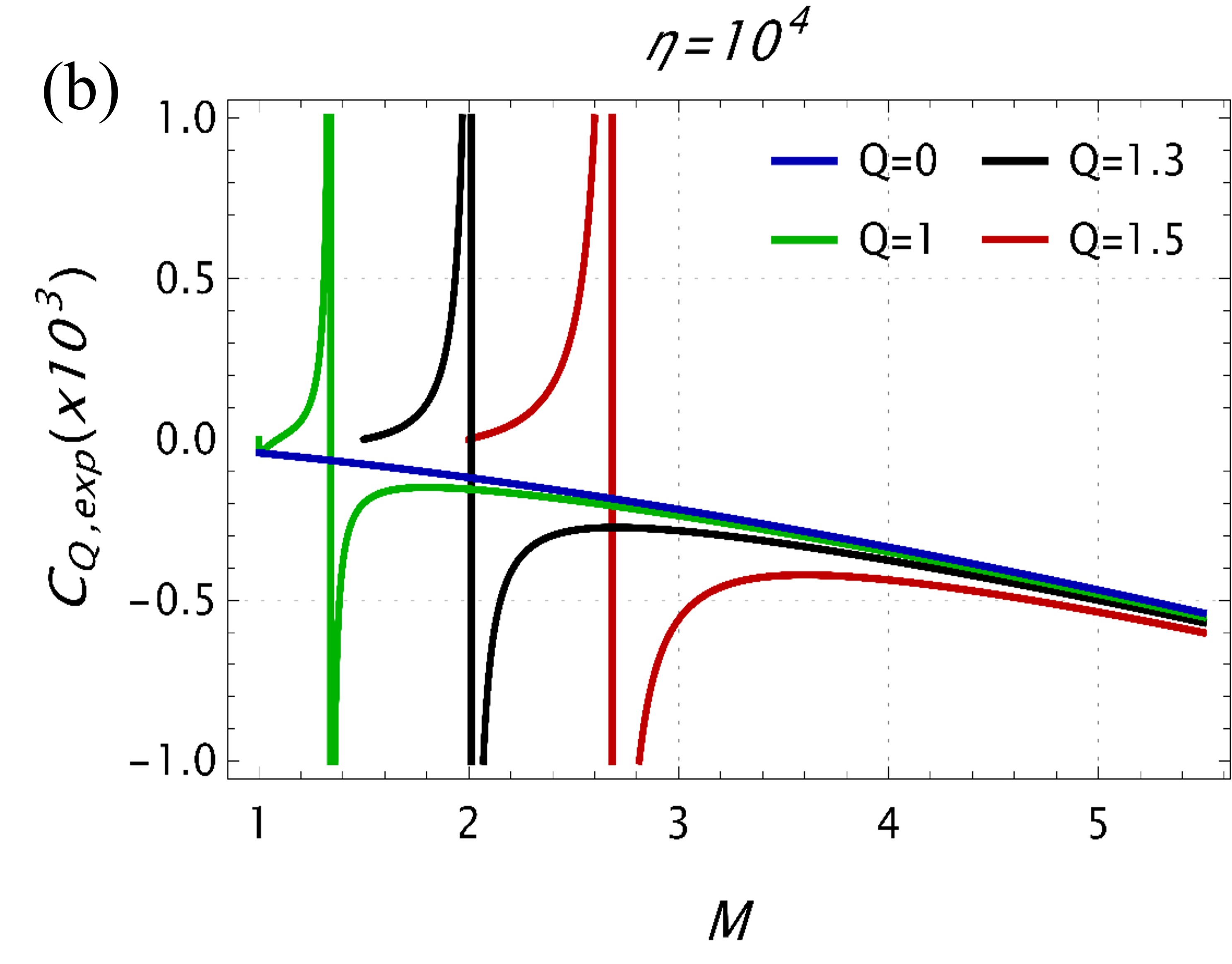}
\caption{ (a) Impact of $\eta$: heat capacity $C_{Q,exp}$ (linear scale) vs BH mass $M$ (log scale), and (b) Impact of charge $Q$ in presence of $\eta$.}
\label{Cexp}
\end{figure*}

For exponential corrections, we use  eq. (\ref{ourentropy}) to first compute 
\begin{align}\nonumber
 \frac{\partial S_{exp}}{\partial M} &=\frac{1}{4 \sqrt{(M-Q) (M+Q)}}\Bigg[3 \pi ^2 e^{-\frac{1}{2} \pi ^2 \left(\sqrt{M^2-Q^2}+M\right)^{3/2}}\\
\label{partialS}
 & \times \left(\sqrt{M^2-Q^2}+M\right)^{3/2}
  \left(e^{\frac{1}{2} \pi ^2
   \left(\sqrt{M^2-Q^2}+M\right)^{3/2}}-\eta \right)\Bigg],
 \end{align}
 and
 \vspace{-10pt}
\begin{eqnarray}\label{partialT}
\frac{\partial T}{\partial M}=\frac{6 Q^2-2 M \left(\sqrt{(M-Q) (M+Q)}+M\right)}{\sqrt{(M-Q) (M+Q)} \left(\sqrt{(M-Q) (M+Q)}+M\right)^{5/2}}.
\end{eqnarray}
 \vspace{-10pt}
Substituting Eqs. \ref{temp1}, \ref{partialS} and \ref{partialT} in eq. (\ref{CQ}) yields
\begin{widetext}
\begin{align}\nonumber
C_{Q,exp}&=\frac{1}{2 \left(M \left(\sqrt{(M-Q) (M+Q)}+M\right)-3 Q^2\right)} \Bigg[ 3 \pi ^2 e^{-\frac{1}{2} \pi ^2 \left(\sqrt{(M-Q) (M+Q)}+M\right)^{3/2}}\times \left(\sqrt{(M-Q) (M+Q)}+M\right)^{3/2} \\
\label{CQ5D}
& \times \left(M \left(\sqrt{(M-Q) (M+Q)}+M\right)-Q^2\right)  \left(e^{\frac{1}{2} \pi ^2 \left(\sqrt{M^2-Q^2}+M\right)^{3/2}}-\eta \right)\Bigg],
\end{align}
\end{widetext}
which evidently incorporates the non-perturbative corrections parameterized by $\eta$. By plotting $C_{Q,exp}$ in figure \ref{Cexp},  we perform a graphical analysis to infer what happens to the thermodynamic behavior of our BH as its size shrinks.   

The first thing we observe here is that, in  both uncorrected ($\eta=0$) and corrected ($\eta\neq 0$) cases, $C_{Q,exp}$ stays negative for larger BH sizes, and  suffers from an infinite discontinuity as it shrinks further in size. At this point, it turns from negative to positive, thereby manifesting an unstable to stable phase transition.    This change, representing a \textit{second order phase transition},  is a feature somewhat peculiar to charged BHs, conceptualized by Davies \cite{1978RPPh...41.1313D}, and is purely of geometric origin due to the presence of horizons in the spacetime. Normally, for Schwarzschild BHs, heat capacity always remains negative, pointing toward their thermodynamic instability. Adding charge to the Schwarzschild BH lends a kind of stability to it. Consider, from the Davies' point of view,  the heat capacity for a $4$D charged BH given by \cite{1978RPPh...41.1313D}
\begin{eqnarray}\label{davies}
    C_{Q,4D}=\frac{8MS^3T}{1/4Q^2-8T^2S^3},
\end{eqnarray}
where $Q$ and $M$ are as usual BH charge and mass, respectively. Setting $Q=0$ makes $C_{Q,4D}$ negative which represents an unstable Schwarzschild BH. Further, one notes that the limit $T\rightarrow 0$ makes the $C_{Q,4D}\rightarrow 0$ via positive values. One concludes that for a certain value of $Q$, $C_{Q,4D}$ must have changed sign. This occurs when the denominator of Eq. (\ref{davies}) vanishes. In our case, we face similar situation where for some particular value of $Q$, $C_{Q}$, given by Eq. (\ref{CQ5D}), changes sign through an infinite discontinuity. These unstable to stable phase transitions classify as the second order phase transitions, quite ubiquitous in nature. The very familiar ferromagnetic to paramagnetic, conductor to superconductor, or liquid-crystal phase transitions are the prime examples of second order phase transition which bear close similarity to our case.

Once we enter  quantum domain,  $C_{Q}$ then again turns negative through zero in presence of $\eta$, and tends to be more negative (unstable) for larger $\eta$. It finally goes to zero at extremal limit ($M=Q$). This negativity of $C_{Q}$ only occurs in presence of $\eta$, and is absent for $\eta=0$ case. Hence, $\eta$ lends different behaviours to the end stages of our BH as it approaches extremal geometry. $C_{Q}=0$ signifies a \textit{first order phase transition}.  We thus conclude that, with $eta$, our BH always remains thermodynamically in an unstable phase for larger sizes and attains stability for some region before again becoming unstable. So for classical geometries, our BH is unstable, and it undergoes stable/unstable phase transitions in quantum regime. Roots of $C_{Q}$ indicate, what generally are known as bound points, which separate physically acceptable positive temperature solutions from  negative (unphysical) ones \cite{Hendi:2018sbe}. However, in our case, in addition to temperature considerations, it aids in identifying  a critical mass $M_{c}$, corresponding to the first root of $C_{Q,exp}$, which marks the onset of  phase transitions. We find $M_{c}$ to be
\begin{widetext}
\begin{align}
M_{c}=\frac{\left(16 \log ^8(\eta )+\pi ^8 Q^6 \log ^4(\eta )+\sqrt{\log ^8(\eta ) \left(\pi ^8 Q^6-16 \log ^4(\eta )\right)^2}\right)^{2/3}+4 \pi ^{8/3} Q^2 \log ^4(\eta )}{4 \pi ^{4/3}
   \log ^2(\eta ) \sqrt[3]{16 \log ^8(\eta )+\pi ^8 Q^6 \log ^4(\eta )+\sqrt{\log ^8(\eta ) \left(\pi ^8 Q^6-16 \log ^4(\eta )\right)^2}}},
\end{align}
\end{widetext}
which obviously depends on  $\eta$ and $Q$ on the BH, and this phase transition is absent in original $5$D RNBHs, and has its sole origin in non-perturbative corrections to entropy. Since non-perturbative corrections become relevant only at small (quantum) scales, it is appropriate to treat this as a  large (classical) to small (quantum) BH phase transition, quite ubiquitous in  BHs \cite{Kubiznak:2016qmn}. The second zero of heat capacity characterizes a BH that does not exchange energy with its surroundings.  This means that the BH ceases evaporation at this stage and ends up as a black remnant. It is hard to ascertain purely from $C_{Q,exp}$ alone what happens beyond this point since our study only concerns up to the extremal limit.  
As we will see later,  thermodynamic geometry conveys much richer structure than heat capacity at extremal limit.   This situation finds its parallel in the role of $Q$ as depicted in figure \ref{Cexp} (b). The larger the $Q$, evaporation stops at a larger $M$, $Q$ just quickens the second order phase transition. 

It is noteworthy from the  right half of the plot in figure \ref{Cexp}  (a) that no matter how big the $\eta$ parameter, all plots overlap and become indiscriminate, which depicts that non-perturbative corrections have no role for macroscopic BHs.  As a side remark, we juxtapose our thermodynamic observation with the gravitational instabilities of five dimensional Reissner-Nordstr\"om BHs. It has been extensively argued  that higher dimensional Reissner-Nordstr\"om BHs in D$<7$ dimensions generally remain gravitationally stable against large values of  $Q$ \cite{Kodama:2003ck, PhysRevLett.103.161101}. However, from thermodynamic point of view,   our BH shows an unstable phase for larger sizes as well as smaller sizes. Our findings conform to the arguments presented in  Refs. \cite{Kodama:2003ck, PhysRevLett.103.161101,Huang:2021jaz} for  smaller and larger sizes on either side of discontinuity in $C_{Q,exp}$ as seen from figure \ref{Cexp} (b), however, it does not so for larger radii. It would be interesting to take this correlation further, which would perhaps span a separate work. 

\subsubsection{Logarithmic corrections}  
\begin{figure*}[btp]
\centering
 \includegraphics[height=13.5cm,width=\linewidth]{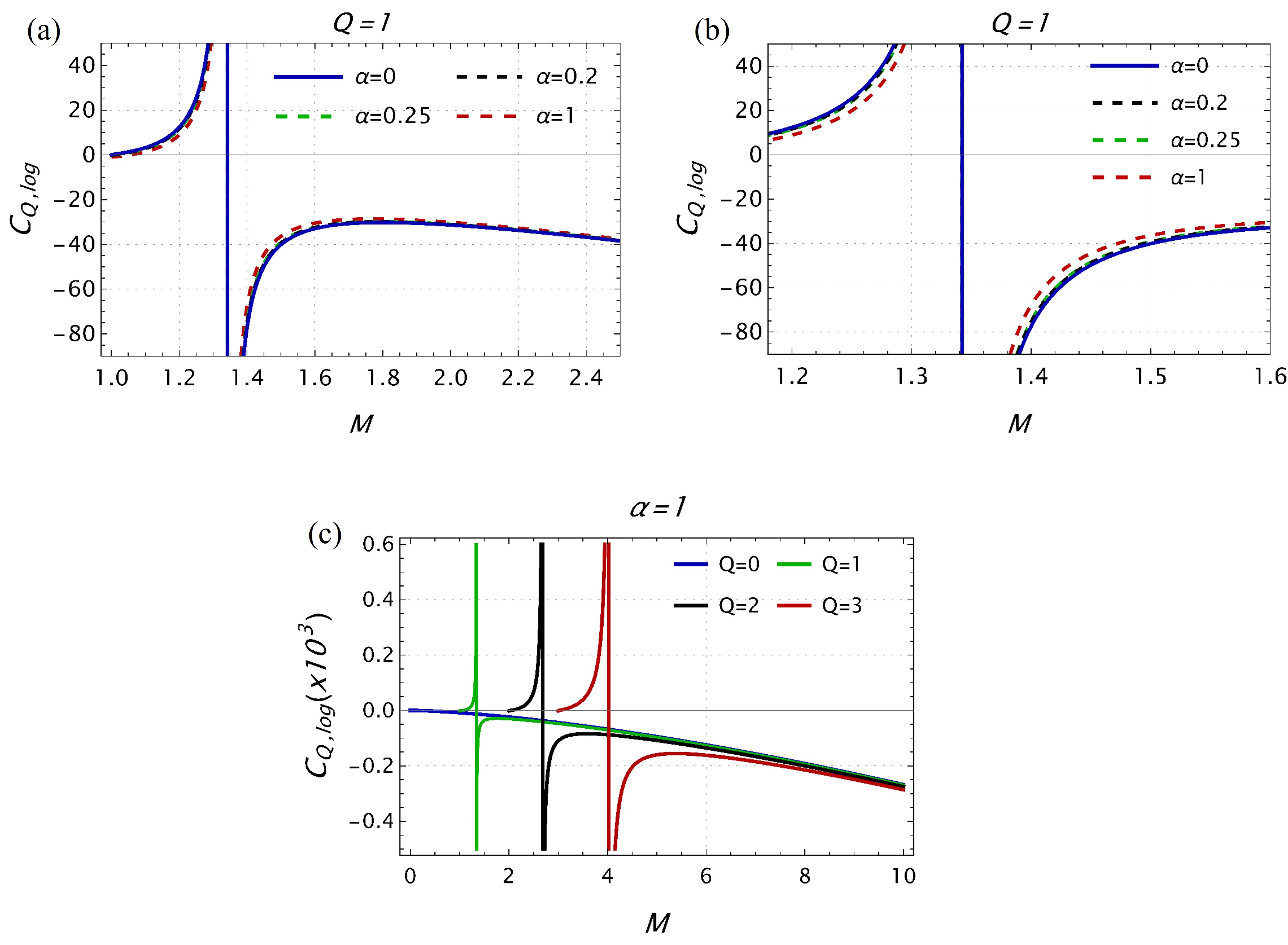}
\caption{Heat capacity $C_{Q,log}$ variation with mass $M$ for logarithmic corrections: (a) Impact of $\alpha$ for fixed $Q$, (b) zoomed-in view of (a), and (c) Impact of $Q$ for a fixed $\alpha$.}
\label{Clog}
\end{figure*}
In this case, the heat capacity is given by
\begin{widetext}
\begin{align}\nonumber
 C_{Q,log}&=-\frac{\left(M \left(\sqrt{(M-Q) (M+Q)}+M\right)-Q^2\right)}{2 \sqrt{(M-Q) (M+Q)} \left(M \left(\sqrt{(M-Q) (M+Q)}+M\right)-3 Q^2\right)}\Bigg[6 M^2 \sqrt{\sqrt{(M-Q) (M+Q)}+M}-4 \alpha  M -6 Q^2\\
 \nonumber &  \times \left(\sqrt{\sqrt{(M-Q) (M+Q)}+M}\right)+3 \alpha  \sqrt{(M-Q)(M+Q)}
 +6 M \sqrt{(M-Q) (M+Q)} \sqrt{\sqrt{(M-Q) (M+Q)}+M}\Bigg],
\end{align}
\end{widetext}
and plotted in figure \ref{Clog}.

We observe from figure \ref{Clog} that for all cases, our BH possesses negative heat capacity $C_{Q,log}$ (unstable) for larger sizes, and a particular behaviour for the case $\alpha$ is only manifested as one approaches extremal limit $(M=Q)$. Uncorrected case  $\alpha=0$ (blue line) approaches zero and hence BH remains in stable phase  till the remnant forms at $M=Q$.   For $\alpha\neq 0$, with  fixed $Q$ shown in figure \ref{Clog} (a), the infinite discontinuity where the system turns from unstable to stable phase and  which signifies a second order phase transition, occurs at same value of $M$ for all cases. Hence, in general, our BH is unstable and becomes stable before ending up as remnant at $M=Q$. figure \ref{Clog} (b) is the close-up view of figure \ref{Clog} (a), and one can see that for larger values of $\alpha$, heat capacity has increasing trend before the infinite discontinuity, i.e. it tends to make the BH stable. For positive $C_{Q,log}$, after the discontinuity, it lowers heat capacity. Hence it seems thermal fluctuations, embodied in $\alpha$ tend to stabilize BH for large sizes and destabilize it for smaller sizes. The underlying reason may be that for smaller sizes, thermal fluctuations in presence of $\alpha$  make geometry unstable. The infinite discontinuity point is however shifted towards higher $M$ for different values of $Q$ as shown in figure \ref{Clog} (c), which signifies competition between $M$ and $Q$. Note that unlike exponential case, logarithmic modifications do not have a critical mass in $C_{Q,log}$ inasmuch as it would indicate a large to small BH phase transition. Rather, $C_{Q,log}$ possesses a zero only at $M=Q$, which however represents a remnant. In that sense, the critical mass would correspond to the magnitude of $Q$.

\section{\label{sec:Ruppgeometry} Thermodynamic Ruppeiner geometry}
Geometric ideas, as enshrined in thermodynamic geometry, have tremendously advanced our understanding of the thermodynamic structure of black holes. A scalar curvature (an invariant) defined in this parameter space helps us to gain further insight into the phase transitions and  microscopic structure of black holes. The ideas have been proposed in the context of thermal fluctuation theory, which leads to the thermodynamic Riemannian geometry \cite{RevModPhys.67.605}.  These so-called information geometric approaches are expected to potentially provide lessons about microscopic degrees of freedom for BHs \cite{Ruppeiner:2013yca}.  To put it simply, if  a BH has an associated thermodynamic behaviour just like  ordinary gases or fluids, there must be  underlying \textit{micromolecules} with a typical interaction phenomena. We are fortunate enough that information geometry attempts to furnish a deep insight into this microstructure. The first of its kind was formulated by Weinhold \cite{doi:10.1063/1.431689, doi:10.1063/1.431635}, where a metric defined on the state of equilibrium states with components as Hessian of internal energy. The metric is therefore given by 
\begin{eqnarray}
 g_{\mu\nu}^{W}=\partial_{\mu}\partial_{\nu}M(S,N^i),
\end{eqnarray}
where $M$ is internal energy (in geometrized units $c=1$),
 $S$ is the system's entropy, and $N^{i}$ constitute all other extensive parameters of system like volume, internal energy etc. $\mu,\nu=0,1,2,..$ are dimensions that correspond to different extensive parameters.  This construction gives the following line element 
\begin{eqnarray}
 ds_{W}^2=g_{\mu\nu}^{W}dx^{\mu}dx^{\nu},
\end{eqnarray}
from which one can define the curvature scalar (a Gaussian curvature). Inspired by this, Ruppeiner \cite{PhysRevA.20.1608, RevModPhys.67.605}  introduced entropy $S$ in place of $M$ and derived the line element, and it was found that it provides the information about phase transitions. Since then there have been many attempts to extend this information geometric approach to the BH thermodynamics. A Legendre-invariant metric due to Quevedo  \cite{Quevedo:2006xk, Quevedo:2008xn} attempted to resolve some of issues surrounding Weinhold/Ruppeiner formalism, while a more recent to this row is Hendi-Panahiyan-Eslam-Panah-Momennia (HPEM) metric \cite{Hendi:2015rja}. Here, we employ the formalism due to Ruppeiner to our BH system as it evaporates to smaller sizes and attempt to reveal the underlying  transformation as the hole reduces to quantum scales. Previously, it has been found that all higher-dimensional variants of RNBHs manifest a flat Ruppeiner geometry (zero curvature), thereby indicating an ideal state behaviour \cite{Aman:2005xk}. However, here we show that the case is not so when the hole size approaches quantum regime dominated by perturbative or non-perturbative quantum corrections. The curvature scalar diverges for exponential case and indicates a phase transition at smaller sizes, which coincides with the zero of $C_{Q,exp}$ (at extremal limit) reported earlier in Section \ref{sec:thermstability}.

We begin from the well-known Boltzmann entropy relation
\begin{eqnarray}
 S=k_{B}\ln\Omega,
\end{eqnarray}
where $k_{B}$ is Boltzmann constant and $\Omega$ denotes the number of microstates of system. The inversion of $\Omega$
\begin{eqnarray}\label{omega}
 \Omega=\exp{\left(\frac{S}{k_{B}}\right)},
\end{eqnarray}
acts as starting point of thermodynamic fluctuation theory from which the Ruppeiner approach emerges.
Consider a set of parameters $x^{0}$ and $x^1$ which characterize a thermodynamic system (here the BH). The probability of finding this system in the intervals $x^0+dx^0$ and $x^1+dx^1$ is given by 
\begin{eqnarray}
 P(x^0,x^1)dx^0dx^1= \mathcal{A} \Omega(x^0,x^1)dx^0dx^1,
\end{eqnarray}
where $\mathcal{A}$ is a normalization constant.
Upon using eq. (\ref{omega}), we can write 
\begin{eqnarray}
 P(x^0,x^1)\propto \exp{\left(\frac{S}{k_{B}}\right)},
\end{eqnarray}
and 
\begin{eqnarray}
 S(x^0,x^1)=S_{bh}(x^0,x^1)+S_{E}(x^0,x^1),
\end{eqnarray}
where $S_{bh}$ is the BH, $S_{E}$ the environment entropy, such that $S_{bh}<S_{E}\sim S$.  For a small change in entropy around equilibrium point $x_{0}^\mu$ (where $\mu,\nu=0,1$), we can write the total entropy by expanding it around the equilibrium, 
\begin{align}\nonumber
 S&=S_{0}+\frac{\partial S_{bh}}{\partial x^\mu}\bigg|_{x^\mu=x_{0}^\mu}\Delta x_{bh}^\mu+\frac{\partial S_{E}}{\partial x^\mu}\bigg|_{x^\mu=x_{0}^\mu}\Delta x_{E}^\mu\\
\nonumber  & \quad +\frac{1}{2}\frac{\partial ^2 S_{bh}}{\partial x^\mu\partial x^\nu}\bigg|_{x^\mu=x_{0}^\mu}\Delta x_{bh}^{\mu}\Delta x_{bh}^{\nu}+\frac{1}{2}\frac{\partial ^2 S_{E}}{\partial x^\mu\partial x^\nu}\bigg|_{x^\mu=x_{0}^\mu}\Delta x_{E}^{\mu}\Delta x_{E}^{\nu}+....,
\end{align}
where $S_{0}$ is the equilibrium entropy at $x_{0}^\mu$.  Now, if one assumes a closed system where extensive parameters of BH and environment $x_{bh}^\mu$ and $x_{E}^\mu$, respectively, have conservative additive nature, such that  $x_{bh}^\mu+x_{E}^\mu=x_{total}^\mu=constant$, then we can write
\begin{eqnarray}
 \frac{\partial S_{bh}}{\partial x^\mu}\bigg|_{x^\mu=x_{0}^\mu}\Delta x_{bh}^\mu=-\frac{\partial S_{E}}{\partial x^\mu}\bigg|_{x^\mu=x_{0}^\mu}\Delta x_{E}^\mu.
\end{eqnarray}
This leads us to 
\begin{eqnarray}\label{BHE}
 \Delta S=\frac{1}{2}\frac{\partial^2 S_{bh}}{\partial x^\mu \partial x^\nu}\bigg|_{x^\mu=x_{0}^\mu}\Delta x_{bh}^{\mu}\Delta x_{bh}^{\nu}+\frac{1}{2}\frac{\partial^2 S_{E}}{\partial x^\mu \partial x^\nu}\bigg|_{x^\mu=x_{0}^\mu}\Delta x_{E}^{\mu}\Delta x_{E}^{\nu}.
\end{eqnarray}
As $S_{E}\sim S$, the second term in eq. (\ref{BHE}) is very small and can be ignored, which leaves behind only BH system with the probability given by 
\begin{eqnarray}
 P(x^0,x^1)\propto \exp{\left(-\frac{1}{2}\Delta l^2\right)},
\end{eqnarray}
where $\Delta l^2$ is given by 
\begin{eqnarray}
 \Delta l^2=-\frac{1}{k_{B}}g_{\mu\nu}\Delta x^{\mu}\Delta x^{\nu}.
\end{eqnarray}
If we set $k_{B}=1$, we get 
\begin{eqnarray}\label{FBO}
 \Delta l^2=g_{\mu\nu}\Delta x^{\mu}\Delta x^{\nu},
\end{eqnarray}
where 
\begin{eqnarray}
 g_{\mu\nu}=-\frac{\partial^2 S_{bh}}{\partial x^\mu \partial x^\nu}.
\end{eqnarray}

In eq. (\ref{FBO}), $\Delta l^2$ is a dimensionless, positive definite, invariant quantity, since probability is a scalar quantity. The  above line element closely resembles the one  in Einstein gravity, and conventionally interpreted as being  the thermodynamic length between two equilibrium fluctuation states: \textit{thermodynamic states are further apart if the fluctuation probability is less} \cite{Ruppeiner:2013yca}. This is in line with the familiar Le Chatelier’s principle that assures a local thermodynamic stability.  The corresponding metric, after dropping the subscript $bh$, reads
\begin{eqnarray}
 g_{\mu\nu}=-\frac{\partial^2 S}{\partial x^\mu \partial x^\nu},
\end{eqnarray}
which is the famous Ruppeiner metric. It is possible to define a curvature scalar for the above line element, similar to what one does in Riemannian geometry. For that matter, consider the Christoffel symbols
\begin{eqnarray}
 \Gamma_{\mu\nu}^\sigma=\frac{1}{2}g^{\sigma\rho}\left(\partial_{\nu}g_{\rho\mu}+\partial_{\mu}g_{\rho\nu}-\partial_{\rho}g_{\mu\nu}\right),
\end{eqnarray}
and the Riemann tensor
\begin{eqnarray}
 R_{\rho\mu\nu}^\sigma=\partial_{\nu}\Gamma_{\rho\mu}^\sigma-\partial_{\mu}
\Gamma_{\rho\nu}^{\sigma}+\Gamma_{\rho\mu}^{\delta}\Gamma_{\delta\nu}^{\sigma}-\Gamma_{\rho\nu}^{\delta}\Gamma_{\delta\mu}^{\sigma},
\end{eqnarray}
from which we define Ricci tensor and scalar as follows 
\begin{eqnarray}
 R_{\mu\nu}=R_{\mu\sigma\nu}^{\sigma}, \ \ R=g^{\mu\nu}R_{\mu\nu}. 
\end{eqnarray}

Applying the above method, one can define curvature scalar for Ruppeiner geometry. It turns out that for a $2$-dimensional space with a  non-diagonal $g_{\mu\nu}$, Ricci curvature scalar reads \cite{Carroll:2004st}
\begin{align}
R&= -\frac{1}{\sqrt{g}}\left[ \frac{\partial}{\partial x^0} \left(\frac{g_{01}}{g_{00}\sqrt{g}}\frac{\partial g_{00}}{\partial x^1}-\frac{1}{\sqrt{g}}\frac{\partial g_{11}}{\partial x^0}\right) \right.\\
 &\left. \ \ \   +\frac{\partial}{\partial x^1}\left(\frac{2}{\sqrt{g}}\frac{\partial g_{01}}{\partial x^0}-\frac{1}{\sqrt{g}}\frac{\partial g_{00}}{\partial x^1}-\frac{g_{01}}{g_{00}\sqrt{g}}\frac{\partial g_{00}}{\partial x^0}\right)\right],
\end{align}
where $g=\det{g_{\mu\nu}}=g_{00}g_{11}-g_{01}^2$.  
The Ruppeiner metric is 
\begin{eqnarray}
 g_{\mu\nu}=-\partial_{\mu}\partial_{\nu}S(M,N^i),
\end{eqnarray}
where $M$ is the BH mass, and $N^i$ the set of other extensive parameters. Naturally for our case, we choose charge $Q$ as second extensive variable. The line element therefore reads
\begin{eqnarray}
 ds_{R}^2=g_{MM}dM^2+2g_{MQ}dMdQ+g_{QQ}dQ^2,
\end{eqnarray}
with the metric 
\[
  g_{\mu\nu} =
  \left( {\begin{array}{cc}
    g_{MM} & g_{MQ} \\
    g_{QM} & g_{QQ} \\
  \end{array} } \right).
\]
The components of $g_{\mu\nu}$ are given by 
\begin{eqnarray}
 g_{MM}=-\frac{\partial}{\partial M}\left(\frac{\partial S}{\partial M}\right),\ g_{MQ}=-\frac{\partial}{\partial M}\left(\frac{\partial S}{\partial Q}\right),
\end{eqnarray}
\begin{eqnarray}
 g_{QM}=-\frac{\partial}{\partial Q}\left(\frac{\partial S}{\partial M}\right),\ \text{and} \ g_{QQ}=-\frac{\partial}{\partial Q}\left(\frac{\partial S}{\partial Q}\right),
\end{eqnarray}
and are detailed in Appendix.
Curvature  is 
\begin{align}
R&= -\frac{1}{\sqrt{g}}\left[ \frac{\partial}{\partial M} \left(\frac{g_{MQ}}{g_{MM}\sqrt{g}}\frac{\partial g_{MM}}{\partial Q}-\frac{1}{\sqrt{g}}\frac{\partial g_{QQ}}{\partial M}\right) \right.\\
 &\left. \ \ \   +\frac{\partial}{\partial Q}\left(\frac{2}{\sqrt{g}}\frac{\partial g_{MQ}}{\partial M}-\frac{1}{\sqrt{g}}\frac{\partial g_{MM}}{\partial Q}-\frac{g_{MQ}}{g_{MM}\sqrt{g}}\frac{\partial g_{MM}}{\partial M}\right)\right],
\end{align}
where $g=\det{g_{\mu\nu}}=g_{MM}g_{QQ}-g_{MQ}^2$ (See Appendix). 

Before computing Ruppeiner curvature, it is imperative to emphasize the interpretation of it. A zero Ruppeiner curvature has been associated with non-interacting BH molecules- much like an ideal gas. A non-zero Ruppeiner curvature depicts non-vanishing interactions between BH molecules. A negative curvature indicates attractive interactions and vice-versa \cite{Ruppeiner:2013yca}. If that is the case, a negative curvature would allude to the existence of a stable system. Since Ruppeiner curvature signifies interactions between BH constituents, one would expect a BH system always have a large curvature since it is a collapsed object with incredible density. However, following Ruppeiner's reasoning \cite{Ruppeiner:2013yca}, it seems convincing to assume that gravitational degrees of freedom responsible for holding up the system elements might have a non-statistical nature since the gravitating particles have collapsed into central singularity. Thermodynamic curvature merely reflects the interactions (perhaps non-gravitational) among the fluctuating thermodynamic constituents at BH surface originating from the underlying gravity-bound system. In that case, associating an ideal gas-like  behaviour with zero curvature makes perfect sense.\\
\indent \textcolor{black}{The usual picture of Ruppeiner geometry considers a classical geometry, quantum corrections to geometry and entropy change things significantly on small scales. In fact, on quantum scales, it is a kind of non-equilibrium description of BH thermodynamics \cite{Pourhassan:2021mhb,Pourhassan:2022irk,Iso:2011gb}. Since present work considers quantum corrections to the thermodynamics (and not the geometry), our definition of Ruppeiner geometry is also somehow effective as it includes a quantum-corrected thermodynamic entropy. Thus, it is reasonable to think that this formulation of Ruppeiner geometry is extrapolated to quantum scales. Such work has been done in Ref. \cite{Pourhassan:2020bzu} which considers a Born-Infeld BH.} 

\subsection{Exponential corrections} 

\begin{figure*}[btp!]
\centering
\includegraphics[width=0.9\linewidth,height=12.1cm]{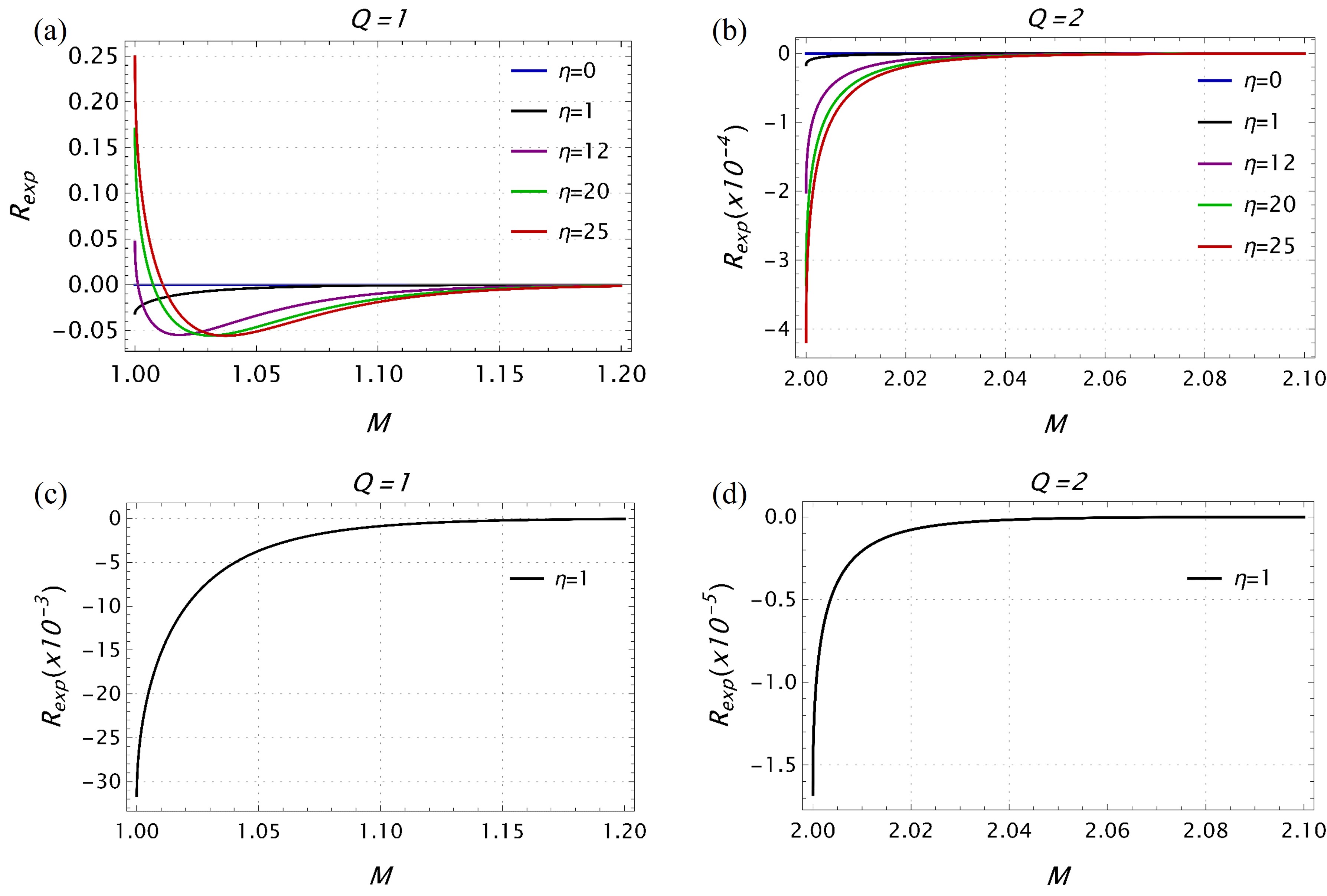}\caption{Impact of $\eta$ on  Ruppeiner curvature scalar $R_{exp}$ for a fixed $Q$.  $R_{exp}$  is zero for bigger hole sizes until $M=Q$, where it diverges and shows a phase transition.}%
\label{Rupp2dd}%
\end{figure*}
In this section, we discuss thermodynamic geometry of $5$D RNBH in presence of exponential corrections $(\eta)$. Since the final expression for Ruppeiner curvature $R_{exp}$ turns out to be too long, henceforth we only carry out graphical analysis by plotting $R_{exp}$ for a range of parameters. It is possible  to plot $R_{exp}$ as a function of mass $M$ while keeping $Q$ fixed. Since for non-extremal case, $M$ exceeds $Q$ which means horizon radii is mostly governed by $M$ than $Q$. We quantify the role of $\eta$ and $Q$ separately. To this end, we present a $2$d plot of $R_{exp}$ for two different values of $Q$ in figure \ref{Rupp2dd}.

Figure \ref{Rupp2dd} (a) is for  $Q=1$ and Fig. \ref{Rupp2dd} (a) for the case $Q=2$.  One can see in both cases, for large sizes with bigger $M$, $R_{exp}$ is zero and changes radically as $M\rightarrow Q$, the quantum domain. Thus our BH manifests a flat geometry for larger sizes and becomes curved (negative or positive) while approaching the extremal limit. This in other words indicates an ideal gas like behaviour for larger sizes, while manifesting multiple phase transitions for smaller (quantum) sizes.   At $M=Q$, $R_{exp}\rightarrow \pm \infty$ depending on the choice of $\eta$ and $Q$ signalling a phase transition. First consider the case $Q=1$. As shown in figure \ref{Rupp2dd} (a), for $\eta=1$ (black curve) $\left [\text{see also figure \ref{Rupp2dd} (c) for a clear view} \right ]$, $R_{exp}$ diverges to $-\infty$, whilst rest of the cases show positive divergence. Hence we conclude that for $\eta=1$, our BH ends up in a stable phase and unstable for rest of the cases where  $R_{exp} \rightarrow +\infty$. $\eta=1$ case possesses only two phase transitions while as rest of the cases have more than two.
 The first phase transition is where $R_{exp}$ turns from zero to negative, and second one when it goes to positive through zero, before diverging at $M=Q$. So our BH changes from ideal to stable phase, then again ideal phase (momentarily) before becoming unstable.      
Hence exponential corrections lend a region of stability for the BH before the final phase transition at $M=Q$. Beyond $M=Q$, $R_{exp}$ becomes imaginary and can't tell anything about the system. For the case $Q=2$, we find all curves behave like $\eta=Q=1$, which means in this case, the BH ends up in a stable phase.  

\begin{figure*}[hbtp!]
\centering
\includegraphics[width=\linewidth,height=13.5cm]{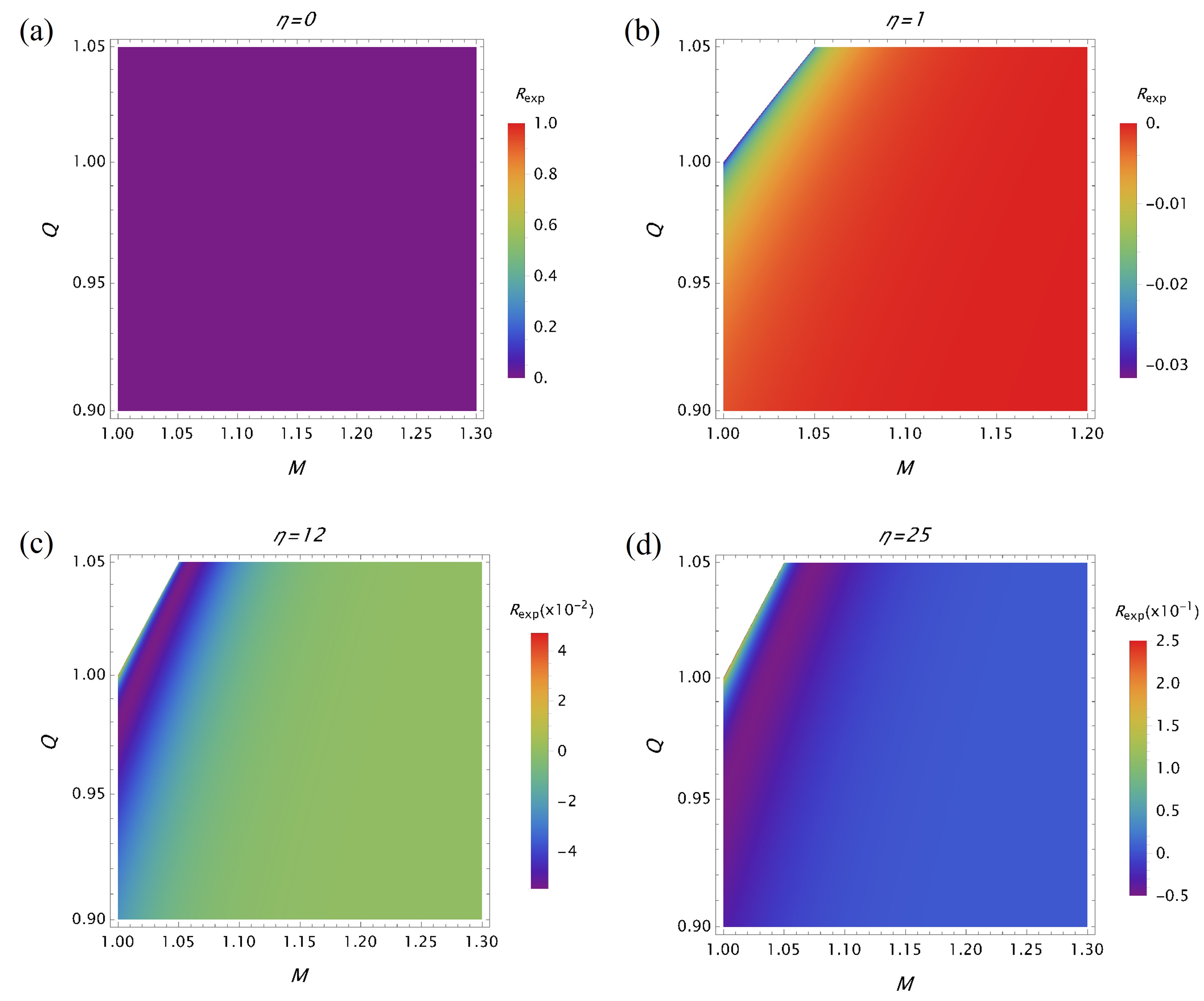}\caption{Density plot of $R_{exp}$ to show different regions of stability/instability. Beyond $M=Q$, $R_{exp}$ is imaginary.}%
\label{RuppcpQ1}%
\end{figure*}
  \begin{figure*}[hbtp!]
\centering
\includegraphics[width=\linewidth,height=13.0cm]{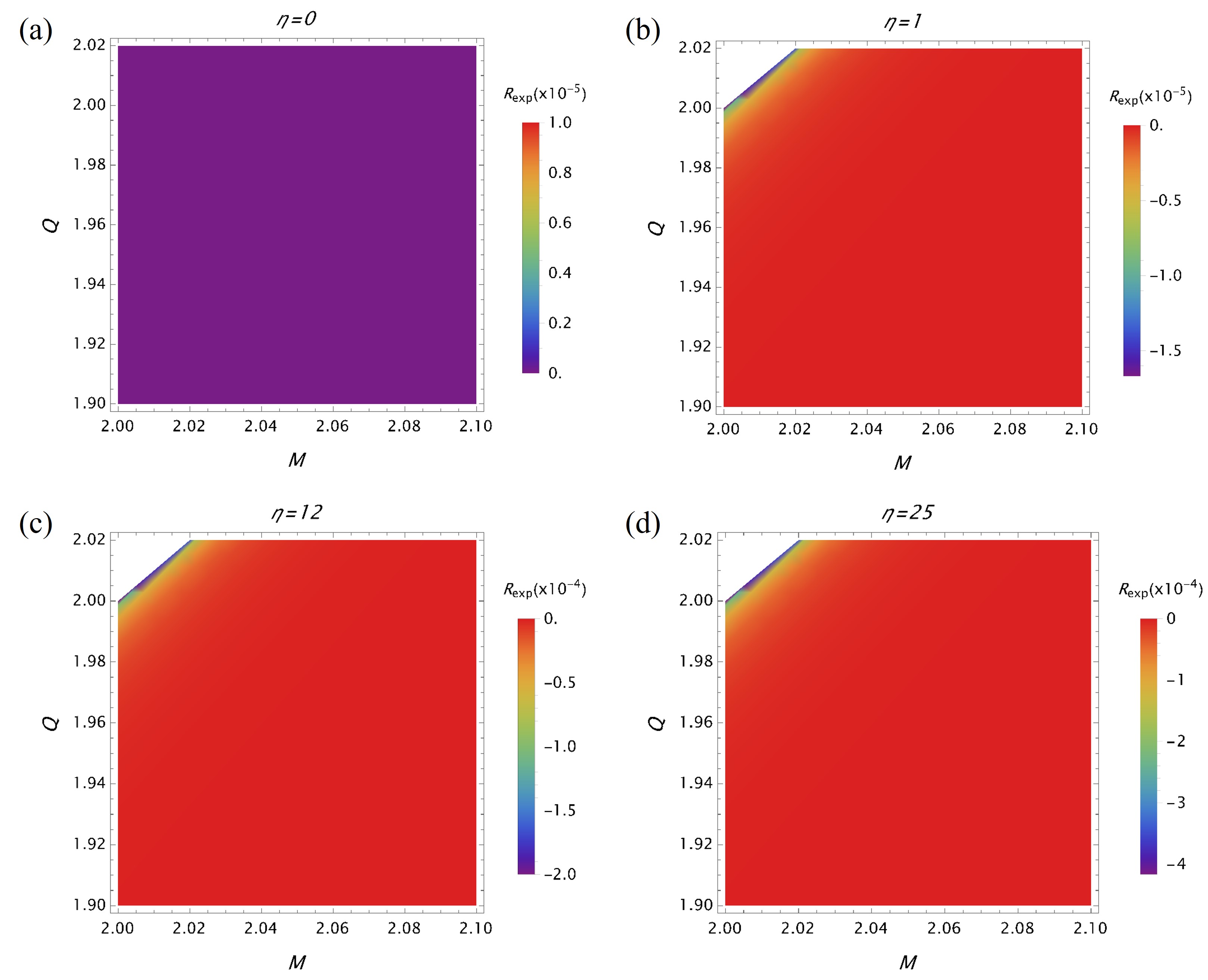}\caption{Density plot of $R_{exp}$ to show different regions of stability/instability for the case $Q=2$.}%
\label{RuppcpQ2}%
\end{figure*}
\begin{figure*}[htb!]
\centering
\includegraphics[height=7cm,width=0.46\textwidth]{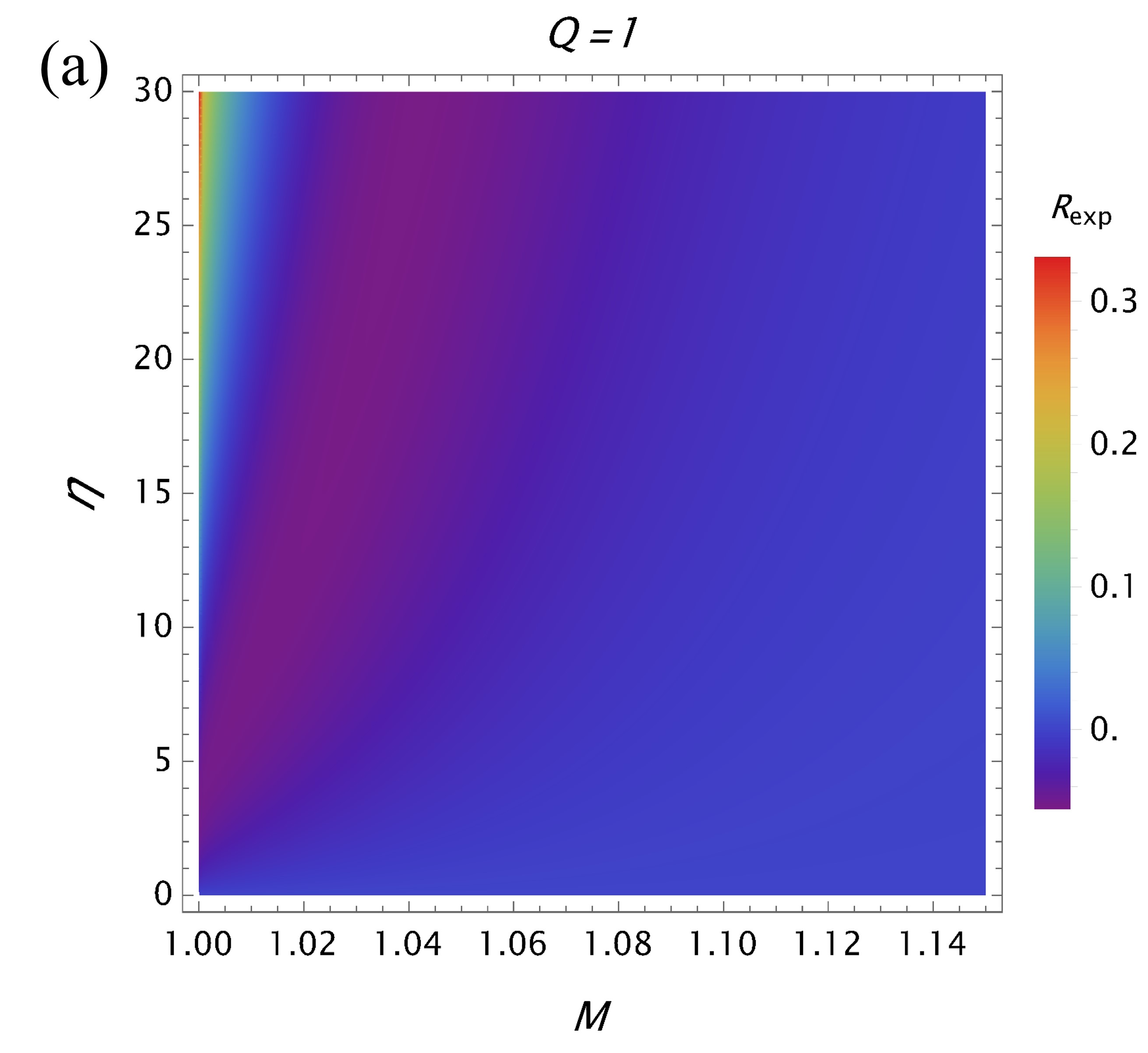}\ \ 
\includegraphics[height=7cm,width=0.46\textwidth]{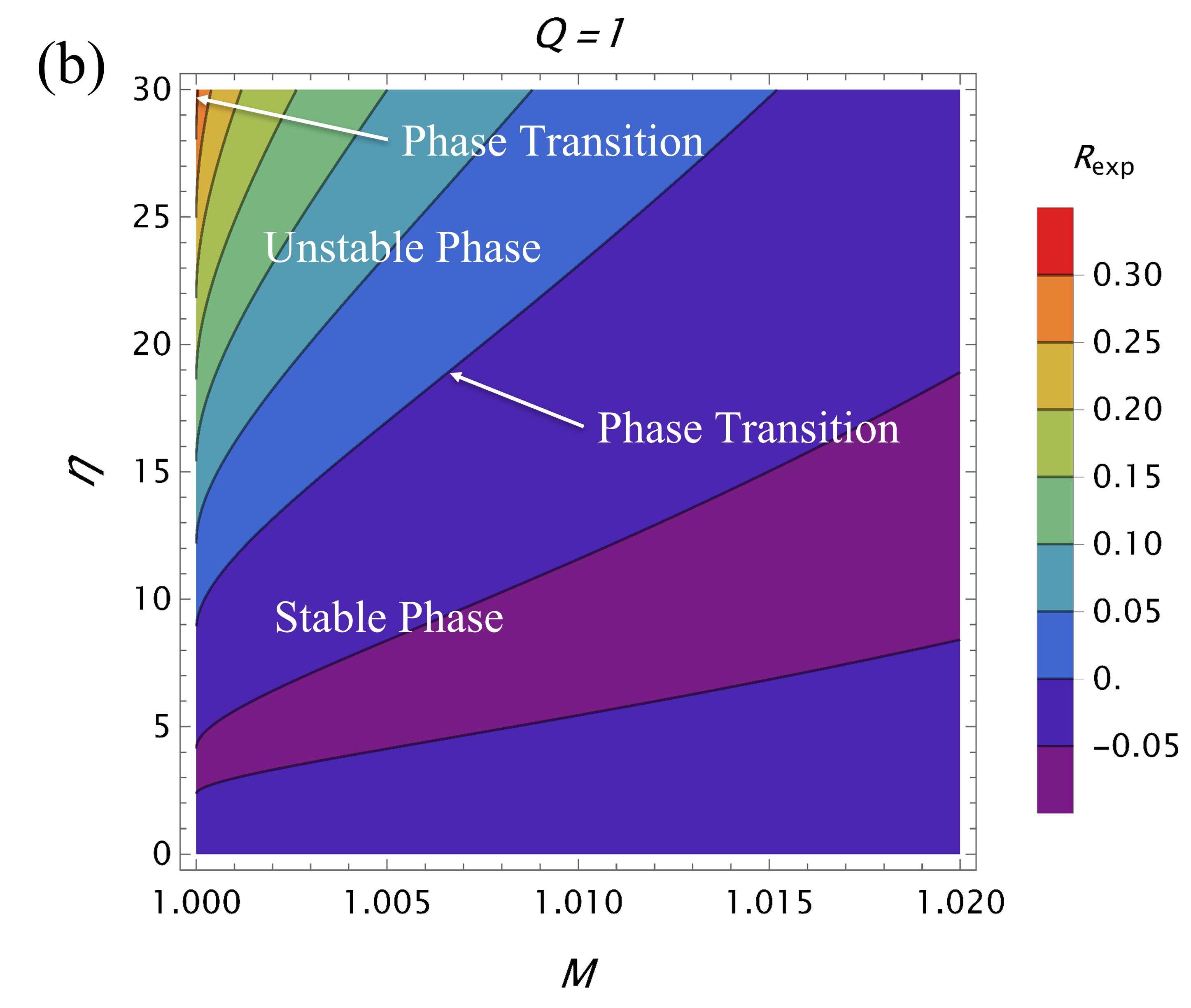}
\caption{Parameter space of $R_{exp}$ with interplay of $M$ and $\eta$ for fixed $Q$: (a) large BH size view, and (b) small BH size view. Multiple phase transitions can be seen here which arise out of $\eta$.}
\label{Rparameter}
\end{figure*}

\textcolor{black}{Now, it is possible to relate the behaviour of $R_{exp}$ computed here to the analysis of heat capacity $C_{Q,exp}$ given in Sec. \ref{sec:thermstability}. It is usually believed that singularities in Ruppeiner curvature may coincide with either zeros or divergences of heat capacity. Here, taking a look at Fig. \ref{Rupp2dd}, one sees that in all cases (irrespective of $\eta$), $R_{exp}$ diverges for $M=Q$. Thus for $Q=1$, the divergence of $R_{exp}$ occurs at $M=1$, which coincides with $C_{exp}=0$ as shown in Fig.\ref{Cexp}. The situation holds for  $Q=2$ as well. Thus, in all situations, our BH shows second order phase transition for all $C_{Q,exp}=0$. This affirms  our choice of Ruppeiner curvature being a suitable diagnostic tool for probing it microstructure.}\\
\indent To better appreciate this scenario, we present our results using density plots in Figs. \ref{RuppcpQ1} and \ref{RuppcpQ2}. One can see that it exactly corroborates to the $2$d case as shown above. To be precise, we can see the divergence in $R_{exp}$, beyond which it becomes imaginary. The original unmodified  case corresponding to $\eta=0$ [Fig. \ref{RuppcpQ1}(a)] for both values of $Q$ shows  a flat curvature. It is also important to mention here that divergence points in $R_{exp}$  match with at least one root of $C_{Q,exp}$, i.e. the extremal limit $M=Q$. We also present a parameter space for $R_{exp}$ in figure \ref{Rparameter} with respect to $M$ and $\eta$. For classical geometries, $R_{exp}$ is flat, as seen from figure \ref{Rparameter} (b), whereas the situation changes in quantum regime, where multiple phase transitions occur [see figure \ref{Rparameter} (a)].   
It has been  previously shown in Ref. \cite{Aman:2005xk} that $5$D RNBHs possess flat Ruppeiner geometry for all sizes. Our findings show that this holds true for only classical geometries with original Bekenstein-Hawking entropy. Once quantum gravity-inspired entropy is invoked, these results no longer hold for smaller BH sizes, and most importantly, we rather have multiple phase transitions on small scales.  \\
\indent \textcolor{black}{Our results here are primarily focused on $5$D RNBH. It is possible and perhaps interesting to check the corresponding situation in $4$D case so as to clearly appreciate the role of extra dimensions in dictating the phase structure of BHs. Following the same computational procedure, we thus present a graphical plot of a $4$D RNBH below in Fig.\ref{4D} without writing down the detailed calculations for the sake of brevity. }
\begin{figure}[h]
\centering
\includegraphics[height=6.6cm,width=8cm]{R4D.jpg}
\caption{\label{4D} Ruppeiner curvature $R_{4D}$ of a $4$D RNBH under the impact of $\eta$. Clearly, in contrast to $5$D case, only unstable region exists on small scales.}
\end{figure}
\textcolor{black}{As seen from Fig\ref{4D}, Ruppeiner geometry $R_{4D}$ is flat for larger BH sizes and becomes positive on small scales before turning to zero at extremal limit. Though amplitude increases with $\eta$, however, one infers that irrespective of the choice of the magnitude of $\eta$, our BH remains in unstable phase for all $\eta$ until extremal limit. The situation is different in $5$D case where the choice of $\eta$ decides the positivity or negativity of the curvature. Hence, $4$D RNBH will always have a region of instability in quantum regime before extremal limit. Pertinent to mention that previous study of $4$D RNBH \cite{Aman:2003ug} has indicated  a flat Ruppeiner geometry, exactly same as that of $5$D case. Our results conform to those findings for $\eta=0$. See solid blue curve in Fig.\ref{4D}. Therefore, it is perhaps plausible to conclude that higher-dimensional charged BHs behave thermodynamically very differently on small scales compared to the ordinary $4$D ones, and this disparity stems out of extra dimensions.}

\subsection{Logarithmic corrections}
\begin{figure*}[ht!]
\centering
\includegraphics[height=7.0cm,width=0.48\textwidth]{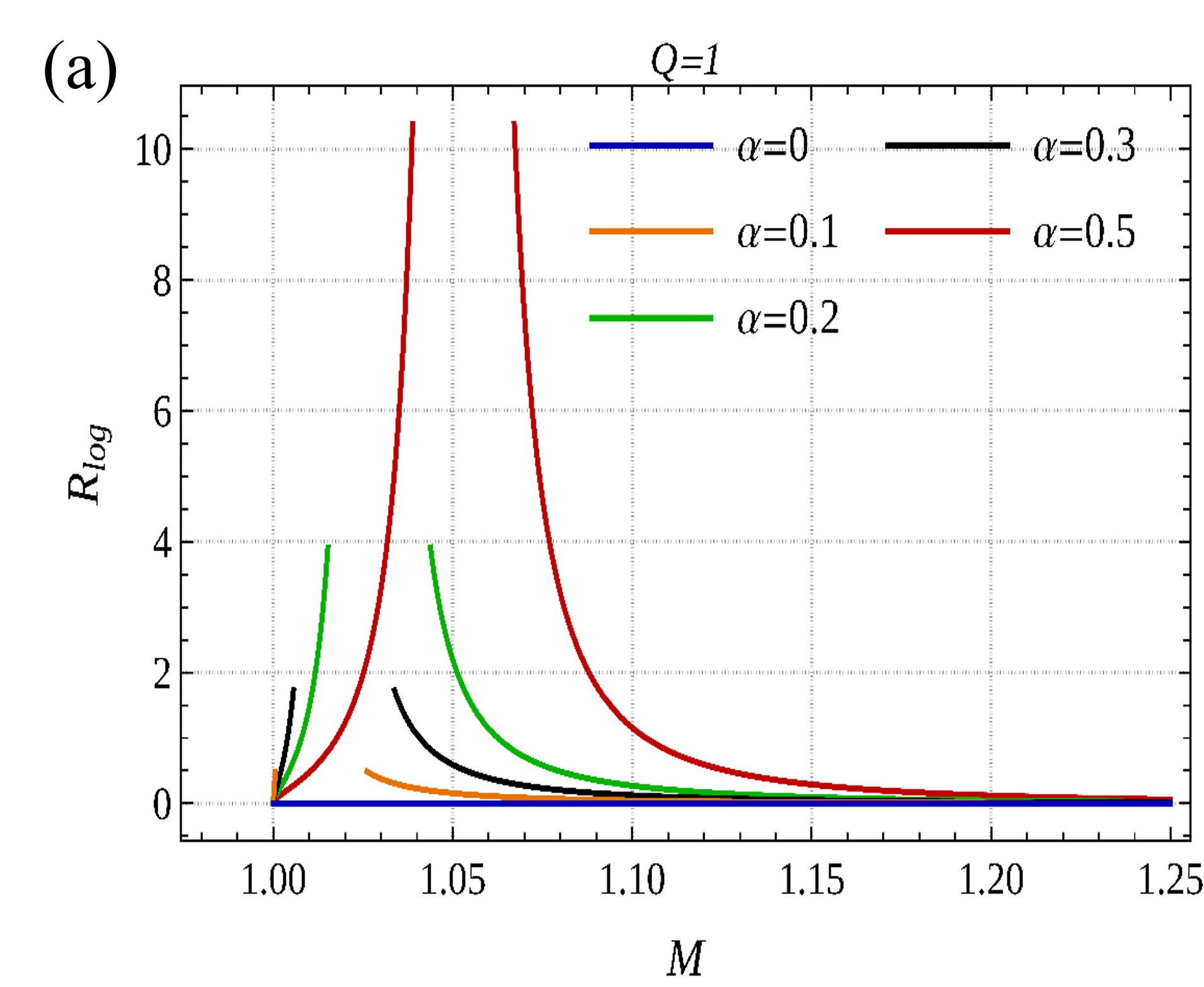}\ \ 
\includegraphics[height=7.0cm,width=0.48\textwidth]{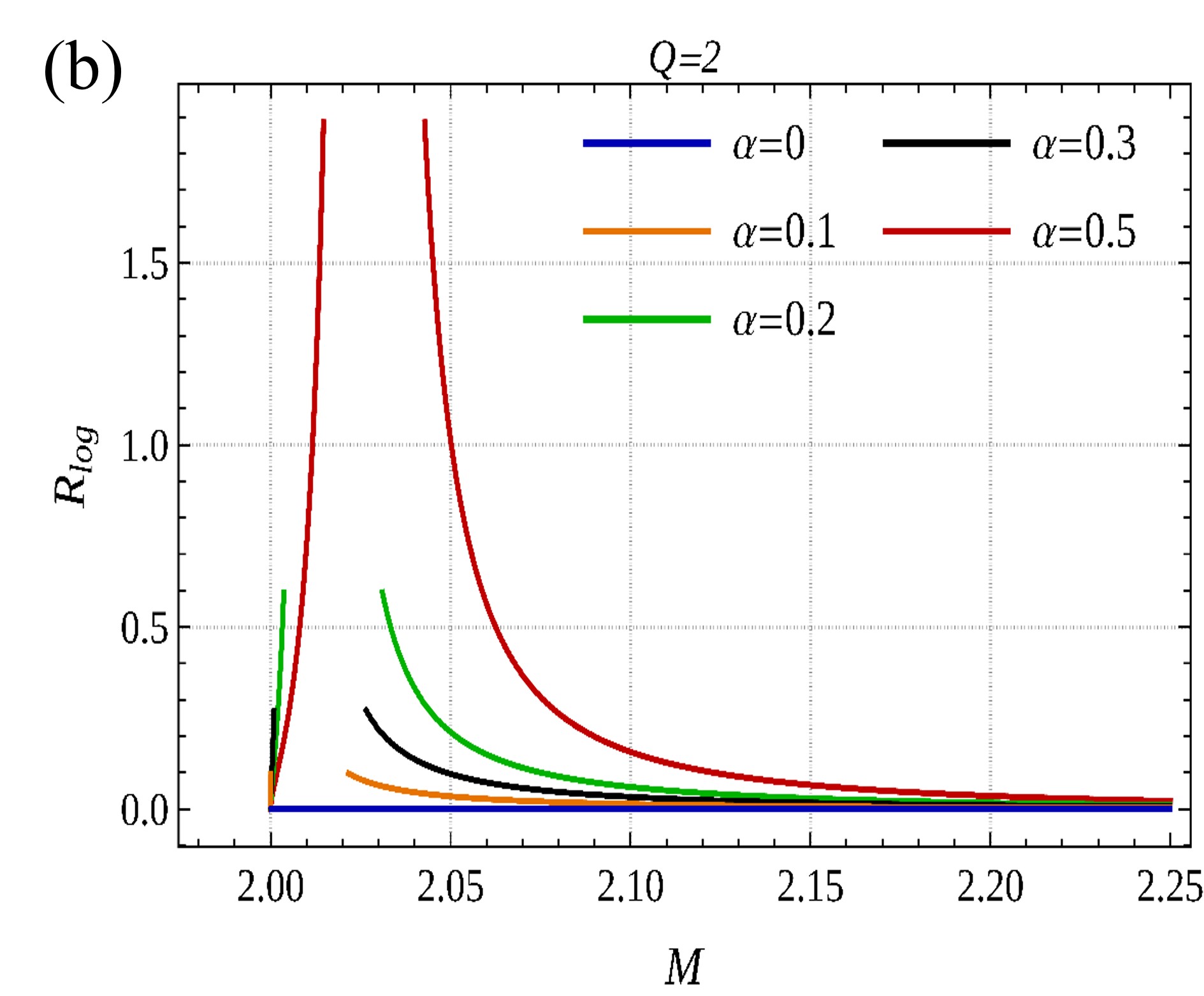}
\caption{Impact of $\alpha$ on  Ruppeiner curvature scalar $R_{log}$.}
\label{Rupplog2d}
\end{figure*}

It would be interesting to check the similar physics of Ruppeiner geometry for logarithmic modifications to BH entropy, which as we said earlier are universal in nature though more pronounced on smaller scales.  
Once again as the expression for curvature turns out to be too long, a graphical analysis would suffice all what we need to unveil. \textcolor{black}{We present the plots in Fig. \ref{Rupplog2d} for two choices of Q}.

Similar to exponential case, Ruppeiner geometry is flat $R_{log}=0$ for larger sizes and become positively curved (unstable) as $M$ decreases. There is a positive divergence, $R_{log}\rightarrow +\infty$, indicating a phase transition, before it goes to zero again (ideal phase). The occurrence of divergence in $R_{log}$ depends on magnitude of logarithmic  corrections $(\alpha)$, with divergence point shifting towards higher $M$ as $\alpha$ increases. Unlike exponential case, there is no correspondence between divergences in $R_{log}$ and heat capacity divergences or zeros. We conclude, from Ruppeiner geometry analysis, that thermal fluctuations tend to make $5$D RNBH unstable in quantum regime before extremal limit. 
\section{\label{sec:conclusion} Conclusion}
The semi-classical formulation of thermodynamics for BHs rests on the Bekenstein-Hawking entropy, which is inadequate to provide any clues for microscopic origin of thermodynamics. Since at present, we have no sensible theory of quantum gravity, attempts to address this question of mirostructure has ushered us in many directions.

Thermodynamic Ruppeiner geometry is a robust candidate to investigate the microstructure of BHs. A curvature defined on the thermodynamic state space of the system tells us about the underlying interactions among BH constituents.  In particular, a positive Ruppeiner curvature shows an unstable system and vice-versa, where as zero curvature indicates an ideal gas-like state. 

Here, we used Ruppeiner geometry to uncover the thermodynamic behaviour of an evaporating $5$D RNBH for both classical and quantum domains, when its entropy is modified by non-perturbative (exponential) and perturbative (logarithmic) contributions.  Our findings suggest that our BH, under the influence of corrections, may undergo several phase transitions as it approaches extremal limit, where mass  and charge  balance each other. For exponential corrections, characterized by $\eta$, whether the system is stable or unstable in the region near and at the extremal point solely depends on the choice of $Q$ and $\eta$. The first phase transition occurs around a critical mass scale which differentiates ideal phase from a stable phase ( $R_{exp}=-\text{ve}$ region).  $R_{exp}$ finally blows up positively (going via zero) or negatively at extremal limit. \textcolor{black}{The divergences of $R_{exp}$ coincide with the zeros of heat capacity $C_{exp}$ for any choice of $\eta$ or $Q$.} For logarithmic modifications quantified by $\alpha$, Ruppeiner curvature $R_{log}$ diverges positively before extremal limit while becoming zero at extremal limit. The divergence point is shifted to larger sizes as $\alpha$ increases. \textcolor{black}{However, unlike exponential case, there exists no correlation between divergences of $R_{log}$ and zeros of heat capacity $C_{Q,log}$. We thus conclude from the above that Ruppeiner geometry might be a better tool to understand the thermal behavior of $5$D RNBHs when the entropy receives exponential corrections. For logarithmic case, it might not be useful, and one should rather look for other formulations of thermodynamic metrics like Weinhold, Quevedo or HPEM metrics. This would be a prospect for future studies.}

We emphasize here that, in absence of quantum gravity modifications, the BH manifests zero curvature (Ruppeiner flat), completely agreeing with previous results that show flat Ruppeiner geometry for RNBHs for all higher spacetime dimensions \cite{Aman:2005xk}.
\section*{Appendix : Computing $g_{\mu\nu}$ for Ruppeiner curvature}
\subsection{Exponential corrections}
The components of metric $g_{\mu\nu}$ are given by
\begin{widetext}
\begin{align*}\nonumber
    g_{MM} &= \frac{-3\pi^2}{16\left(M^2-Q^2\right)^{3/2}}\Bigg[2\sqrt{M+\sqrt{M^2-Q^2}}\bigg\{-3Q^2+M\left(M+\sqrt{M^2-Q^2}\right)\bigg\} +\eta e^{-\frac{\pi^2}{2}\left(M+\sqrt{M^2-Q^2}\right)^{3/2}}\\
\nonumber    &\times \Bigg\{12\pi^2 M^4+3\pi^2 Q^4+12\pi M^3\sqrt{M^2-Q^2} +6Q^2\sqrt{M+\sqrt{M^2-Q^2}} -M\sqrt{M^2-Q^2}\\
 \nonumber & \times   \left(9\pi^2Q^2+2\sqrt{M+\sqrt{M^2-Q^2}}\right)-M^2\left(15\pi^2Q^2+2\sqrt{M+\sqrt{M^2-Q^2}}\right)\Bigg\}\Bigg],\\
g_{MQ}&= \frac{-3\pi^2Qe^{-\frac{\pi^2}{2}\left(M+\sqrt{M^2-Q^2}\right)^{3/2}}}{16\left(M^2-Q^2\right)^{3/2}\sqrt{M+\sqrt{M^2-Q^2}}} \Bigg[2 e^{\frac{\pi^2}{2}\left(M+\sqrt{M^2-Q^2}\right)^{3/2}}\bigg\{-Q^2+M\left(M+\sqrt{M^2-Q^2}\right)\bigg\}\\
    &-\eta\Bigg\{6\pi^2 M^3\sqrt{M+\sqrt{M^2-Q^2}}+2M\bigg(\sqrt{M^2-Q^2}-3\pi^2Q^2\sqrt{M+\sqrt{M^2-Q^2}}\bigg) +Q^2\bigg(2-3\pi^2\sqrt{M^2-Q^2}\\
\nonumber  & \times \sqrt{M+\sqrt{M^2-Q^2}}\bigg)+M^2\bigg(2+6\pi^2\sqrt{M^2-Q^2}
  \times \sqrt{M+\sqrt{M^2-Q^2}}\bigg)\Bigg\}\Bigg],\\
  g_{QM}&= \frac{-3\pi^2Q}{16\left(M^2-Q^2\right)^{3/2}}
\Bigg[2 \left(-2M+\sqrt{M^2-Q^2}\right) \sqrt{M+\sqrt{M^2-Q^2}}+\eta e^{-\frac{\pi^2}{2}\left(M+\sqrt{M^2-Q^2}\right)^{3/2}}\\
& \times \Bigg\{-6\pi^2 M^3\sqrt{M^2-Q^2}+M\bigg(6\pi^2Q^2-4\sqrt{M+\sqrt{M^2-Q^2}}\bigg)+\sqrt{M^2-Q^2}\left(2-3\pi^2\sqrt{M^2-Q^2}\sqrt{M+\sqrt{M^2-Q^2}}\right)
\Bigg\} \Bigg],\\
 g_{QQ}&=\frac{-3\pi^2e^{-\frac{\pi^2}{2}\left(M+\sqrt{M^2-Q^2}\right)^{3/2}}}{16\left(M^2-Q^2\right)^{3/2}\sqrt{M+\sqrt{M^2-Q^2}}}
\Bigg[ e^{\frac{\pi^2}{2}\left(M+\sqrt{M^2-Q^2}\right)^{3/2}}
 \left(-4M^2-4M^2\sqrt{M^2-Q^2}+2Q^2\sqrt{M^2-Q^2}\right)\\
&   +\eta \Bigg\{4M^3-2Q^2\sqrt{M^2-Q^2}-3\pi^2  Q^4\sqrt{M+\sqrt{M^2-Q^2}}
+3\pi^2 M Q^2\sqrt{M^2-Q^2}\sqrt{M+\sqrt{M^2-Q^2}}\\
&+ M^2\left(4\sqrt{M^2-Q^2}+3\pi^2 Q^2\sqrt{M+\sqrt{M^2-Q^2}}\right)
\Bigg\} \Bigg],
\end{align*}
with the determinant
\begin{align*}
g&= \frac{-9\pi^4\left(1-\eta e^{-\frac{\pi^2}{2}\left(M+\sqrt{M^2-Q^2}\right)^{3/2}}\right)}{64\left(M^2-Q^2\right)^2\sqrt{M+\sqrt{M^2-Q^2}}}
 \Bigg[ 2e^{\frac{\pi^2}{2}\left(M+\sqrt{M^2-Q^2}\right)^{3/2}}\sqrt{M+\sqrt{M^2-Q^2}} \left(2M^3-2MQ^2+2M^2\sqrt{M^2-Q^2}-Q^2\sqrt{M^2-Q^2}\right) \\
&  +\eta \Bigg\{24\pi^2 M^5+24\pi^2 M^4\sqrt{M^2-Q^2}+4MQ^2
 \left(3\pi^2Q^2+\sqrt{M+\sqrt{M^2-Q^2}}\right)-4M^2\sqrt{M^2-Q^2}\\
\nonumber
& \times \left(6\pi^2Q^2+\sqrt{M+\sqrt{M^2-Q^2}}\right)+Q^2\sqrt{M^2-Q^2}
\left(3\pi^2+2\sqrt{M+\sqrt{M^2-Q^2}}\right)
\Bigg\} \Bigg].
\end{align*}
\end{widetext}

\subsection{Logarithmic corrections}
The metric elements $g_{\mu\nu}$ read as
\begin{widetext}
\begin{align*}
g_{MM} &= \frac{1}{4 \left(M^2-Q^2\right)^2}\Bigg[-M^2 \left(4 \alpha +3 \sqrt{M^2-Q^2} \sqrt{\sqrt{M^2-Q^2}+M}\right)+3 M \left(\alpha  \sqrt{M^2-Q^2}+Q^2 \sqrt{\sqrt{M^2-Q^2}+M}\right)\\
    \nonumber & +Q^2 \left(9 \sqrt{M^2-Q^2}
   \sqrt{\sqrt{M^2-Q^2}+M}-4 \alpha \right)-3 M^3 \sqrt{\sqrt{M^2-Q^2}+M}\Bigg],\\
g_{MQ} &= -\frac{1}{4\left(M^2-Q^2\right)^2}\Bigg[Q \Bigg(-3 M^2 \sqrt{\sqrt{M^2-Q^2}+M}+6 M \sqrt{(M-Q) (M+Q)} \sqrt{\sqrt{M^2-Q^2}+M}\\
   \nonumber & +3 Q^2 \sqrt{\sqrt{M^2-Q^2}+M}-8 \alpha  M+3 \alpha  \sqrt{(M-Q) (M+Q)}\Bigg)\Bigg],\\
g_{QM}  &= -\frac{1}{4\left(M^2-Q^2\right)^2}\Bigg[Q \Bigg(-3 M^2 \sqrt{\sqrt{M^2-Q^2}+M}+6 M \sqrt{(M-Q) (M+Q)} \sqrt{\sqrt{M^2-Q^2}+M}\\
   \nonumber & +3 Q^2 \sqrt{\sqrt{M^2-Q^2}+M}-8 \alpha  M+3 \alpha  \sqrt{(M-Q) (M+Q)}\Bigg)\Bigg],\\
 g_{QQ} &= \frac{1}{4 \left(Q^3-M^2 Q\right)^2}\Bigg[3 \alpha  M^4-3 M^3 \Big(\alpha  \sqrt{M^2-Q^2}+Q^2 \sqrt{\sqrt{M^2-Q^2}+M}\Big)+M^2 Q^2 \Big(9 \sqrt{M^2-Q^2} \sqrt{\sqrt{M^2-Q^2}+M}-10 \alpha \Big)\\
   \nonumber & +3 M \Big(2\alpha  Q^2 \sqrt{M^2-Q^2}+Q^4 \sqrt{\sqrt{M^2-Q^2}+M}\Big)-Q^4 \Big(\alpha +3 \sqrt{M^2-Q^2} \sqrt{\sqrt{M^2-Q^2}+M}\Big)\Bigg],
    \end{align*}
     with the determinant
    \begin{align*}
     g&= \frac{1}{16 \left(Q^3-M^2 Q\right)^2} \Bigg[-36 M^3 Q^2-3 M^2 \left(7 \alpha ^2+12 Q^2 \sqrt{M^2-Q^2}\right)+\alpha  Q^2 \left(13 \alpha -15 \sqrt{M^2-Q^2} \sqrt{\sqrt{M^2-Q^2}+M}\right)\\
     & +3 M \left(7 \alpha ^2 \sqrt{M^2-Q^2}+8 \alpha  Q^2 \sqrt{\sqrt{M^2-Q^2}+M}+12 Q^4\right)+18 Q^4 \sqrt{M^2-Q^2}\Bigg].
    \end{align*}
\end{widetext}

{\section*{acknowledgments}}
\setlength{\parskip}{0cm}
    \setlength{\parindent}{1em}
SMASB is supported by the CSC Scholarship of China at Zhejiang University.

\bibliographystyle{apsrev4-1}
\bibliography{masood.bib}
\end{document}